\documentclass[11pt]{article}

\usepackage{xspace,amsmath,amssymb,bbm}
\usepackage[sort]{cite}
\usepackage{gastex}
\usepackage{tikz}
\usetikzlibrary{arrows,automata,shapes,calc}
\usepackage[matrix,arrow,cmtip]{xy}
\SelectTips{cm}{}\SilentMatrices\CompileMatrices

\newcommand*\id{\textup{\textsf{id}}}

\newcommand*\E{\mathcal E}

\newcommand*\bbot{%
  \hbox{%
    \hbox to -.475pt{%
      \raisebox{1.5pt}[0pt][0pt]{%
        \rule{6.4pt}{.4pt}%
      }%
      \hss}%
    $\bot$%
  }}
\newcommand*\ttop{%
  \hbox{%
    \hbox to -.475pt{%
      \raisebox{5pt}[0pt][0pt]{%
        \rule{6.4pt}{.4pt}%
      }%
      \hss}%
    $\top$%
  }}
\newcommand*\V{\mathcal V}

\newcommand{\two}{{\mathbf{2}}}
\newcommand{\R}{{\mathcal{R}}}
\newcommand{\FinAdd}{{\mathbf{FinAdd}}}

\newtheorem{theorem}{\bf Theorem}[section]
\newtheorem{lemma}[theorem]{\bf Lemma}        
\newtheorem{proposition}[theorem]{\bf Proposition}  
\newtheorem{corollary}[theorem]{\bf Corollary}
\newtheorem{definition}[theorem]{\bf Definition}

\newtheorem{example}[theorem]{\bf Example}       
\newtheorem{remark}[theorem]{\bf Remark}

\def\Bbox{
{\unskip\nobreak\hfil\penalty50
\hskip1em\hbox{}\nobreak\hfil{\lower .5pt \hbox{$\Box$}}
\parfillskip=0pt \finalhyphendemerits=0 \par}
}

\def\eop{
\ifmmode {\hbox{\Bbox}} \else \Bbox \fi
}

\def\bbox{
\ifmmode {\hbox{\bbox}} \else \Bbox \fi
}

\newcommand*\proof{\textsl{Proof.}\quad}

\newcommand*\Ax[1]{\textsf{Ax#1}}

\begin{document}

\title{{\Large \bf $^*$-Continuous Kleene $\omega$-Algebras}\footnote{The first author acknowledges partial support by grant
  no.~K108448 from the National Foundation of Hungary for Scientific
  Research.  The second and third authors acknowledge partial support
  by ANR MALTHY, grant no.~ANR-13-INSE-0003 from the French National
  Research Foundation.}}

\author{Zolt\'an  \'Esik\\
Dept. of Computer Science\\
University of Szeged\\
Hungary
\and Uli Fahrenberg\\
IRISA / INRIA Rennes\\
France
\and Axel Legay\\
IRISA / INRIA Rennes\\
France
}

\date{\today}

\maketitle

\begin{abstract}
  We define and study basic properties of $^*$-continuous Kleene
  $\omega$-algebras that involve a $^*$-continuous Kleene algebra with
  a $^*$-continuous action on a semimodule and an infinite product
  operation that is also $^*$-continuous. We show that $^*$-continuous
  Kleene $\omega$-algebras give rise to iteration semiring-semimodule
  pairs.  We show how our work can be applied to solve certain energy
  problems for hybrid systems.
\end{abstract}

\section{Introduction}

A \emph{Kleene algebra} \cite{Kozen2} is an idempotent semiring $S = (S,\vee, \cdot, \bot,1)$ 
equipped with a star operation $^*: S \to S$ such that for all $x,y\in S$,
$yx^*$ is the least solution of the fixed point equation $z = zx \vee y$
and 
$x^*y$ is the least solution of the fixed point equation $z = xz \vee y$.

Examples of Kleene algebras include the language semiring $P(A^*) = (P(A^*), \vee, \cdot, \bot, 1)$ 
over an alphabet $A$, whose elements are the subsets of the set $A^*$ of all finite words over $A$, 
and whose operations are set union and concatenation, with the languages 
$\emptyset$ and $\{\epsilon\}$ serving as $\bot$ and $1$. 
Here, $\epsilon$ denotes the empty word. The star 
operation is the usual Kleene star: $X^* = \bigcup_{n \geq 0}X^n = \{u_1\ldots u_n : u_1,\ldots,u_n \in L,\ n \geq 0\}$,
for all $X \subseteq A^*$. 

Another example is the Kleene algebra $P(A \times A) = (P(A \times A),
\vee,\cdot,0,1)$ of binary relations over any set $A$, whose
operations are union, relational composition (written in diagrammatic
order), and where the empty relation $\emptyset$ and the identity
relation $\id$ serve as the constants $\bot$ and $1$. The star
operation is the formation of the reflexive-transitive closure, so
that $R^* = \bigcup_{n \geq 0}R^n$ for all $R\in P(A \times A)$.

The above examples are in fact \emph{continuous Kleene algebras}, i.e., 
idempotent semirings $S$ such that 
equipped with the natural order, they are all complete lattices (hence all
suprema exist),
and the product operation preserves arbitrary suprema in either 
argument:
$$ y(\bigvee X) = \bigvee yX\quad {\rm and}\quad (\bigvee X)y = \bigvee Xy$$
for all $X \subseteq S$ and $y \in S$. 
The star operation is given by $x^* = \bigvee_{n \geq 0} x^n$, 
so that $x^*$ is the supremum of the set $\{x^n : n\geq 0\}$
of all powers of $x$. It is well-known that the language semirings $P(A^*)$ may be identified 
as the \emph{free} continuous Kleene algebras (in a suitable category 
of continuous Kleene algebras). 

A larger class of models is given by the \emph{$^*$-continuous Kleene
  algebras} \cite{Kozen2}. By the definition of $^*$-continuous
Kleene algebra $S = (S,\vee,\cdot,\bot,1)$, only suprema of sets of
the form $\{x^n : n \geq 0\}$ need to exist, where $x$ is any element
of $S$, and $x^*$ is given by this supremum. Moreover, product
preserves such suprema in both of its arguments:
\begin{equation*}
y(\bigvee_{n \geq 0} x^n) = \bigvee_{n \geq 0}yx^n\quad {\rm and}\quad 
(\bigvee_{n \geq 0} x^n)y = \bigvee_{n \geq 0} x^ny.
\end{equation*}
 For any alphabet $A$,
the collection $R(A^*)$ of all regular languages over $A$ 
is an example of a $^*$-continuous Kleene algebra which is 
not a continuous Kleene algebra. The Kleene algebra $R(A^*)$ may be 
identified as the free $^*$-continuous Kleene algebra on $A$. 
It is also the free Kleene algebra on $A$, cf. \cite{Kozen}.
There are several other characterizations of $R(A^*)$, the most 
general of which identifies $R(A^*)$ as the free
iteration semiring on $A$ satisfying the identity $1^* = 1$, cf. \cite{Krob,BEbook}. 

For non-idempotent extensions of the notions of continuous Kleene algebras,
$^*$-continuous Kleene algebras and 
Kleene algebras, we refer to \cite{EsikKuichIND,EsikKuichrational}. 

When $A$ is an alphabet, let $A^\omega$ denote the set of all $\omega$-words 
(sequences) over $A$. An \emph{$\omega$-language} over $A$ is a 
subset of $A^\omega$. It is natural to consider the set $P(A^\omega)$ 
of all languages of $\omega$-words over $A$, equipped with the operation
of set union as $\vee$ and the empty language $\emptyset$ as $\bot$, and the 
left action of $P(A^*)$ on $P(A^\omega)$ defined by $XV = \{xv : x \in X,\ v \in V\}$
for all $X \subseteq A^*$ and $V \subseteq A^\omega$. It is clear that $(P(A^\omega),\vee,\bot)$ 
is in fact a $P(A^*)$-semimodule and thus $(P(A^*), P(A^\omega))$ is a 
semiring-semimodule pair. We may also equip $(P(A^*), P(A^\omega))$ 
with an infinite product operation mapping an $\omega$-sequence
$(X_0,X_1,\ldots)$ over $P(A^*)$ to the $\omega$-language
$\prod_{n \geq 0}X_n = \{x_0x_1\ldots \in A^\omega : x_n \in X_n\}$.
(Thus, an infinite number of the $x_n$ must be different from $\epsilon$.
Note that $1^\omega = \bot$ holds.) 
The semiring-semimodule pair so obtained is a continuous Kleene $\omega$-algebra.

More generally, we call a semiring-semimodule pair $(S,V)$ 
 a \emph{continuous Kleene $\omega$-algebra} if $S$ is a continuous Kleene algebra (hence $S$ and
$V$ are idempotent), $V$ is a complete lattice with the natural 
order, and the action preserves all suprema in either argument. 
Moreover, there is an infinite product operation which 
is compatible with the action and
associative in the sense that the following hold for all 
$x,x_0,x_1,\ldots \in S$: 
\begin{itemize}
\item $x (\prod_{n\geq 0} x_n) = \prod_{n \geq 0}y_n$, where $y_0 = x$ and $y_n = x_{n-1}$ 
for all $n > 0$,
\item $\prod_{n \geq 0} x_n = \prod_{k \geq 0} y_k$, whenever there is a 
sequence of integers $0 = i_0\leq i_1\leq \ldots$ increasing without a bound 
such that $y_k = x_{i_k}\cdots x_{i_{k+1} - 1}$ for all $k \geq 0$. 
\end{itemize}
Moreover, the infinite product operation preserves all suprema:
\begin{itemize}
\item $\prod_{n \geq 0} (\bigvee X_n) = \bigvee \{\prod_{n \geq 0} x_n : x_n \in X_n,\ n \geq 0\}$,
\end{itemize}
for all $X_0,X_1,\ldots \subseteq S$. 

The above notion of continuous Kleene $\omega$-algebra may be seen as a special case of
the not necessarily idempotent complete semiring-semimodule pairs 
of \cite{EsikKuich}. 

Our aim in this paper is to provide an extension of the notion of
continuous Kleene $\omega$-algebras to \emph{$^*$-continuous Kleene
  $\omega$-algebras} which are semiring-semimodule pairs $(S,V)$
consisting of a $^*$-continuous Kleene algebra $S = (S,\vee, \cdot,
\bot,1)$ acting on a necessarily idempotent semimodule $V = (V, \vee,
\bot)$, such that the action preserves certain suprema in its first
argument, and which are equipped with an infinite product operation
satisfying the above compatibility and associativity conditions and
some weaker forms of the last axiom. We will define both a finitary
and a non-finitary version of $^*$-continuous Kleene
$\omega$-algebras.

We will establish several properties of $^*$-continuous Kleene $\omega$-algebras,
including the existence of the supremum of certain subsets related to 
regular $\omega$-languages. Then we will use these results in our 
characterization of the free $^*$-continuous Kleene $\omega$-algebras, at least 
in the finitary version. 

When $(S,V)$ is a $^*$-continuous Kleene algebra, we may define an omega
operation $S \to V$ by $x^\omega = \prod_{n \geq 0} x_n$, where $x_n = x$ 
for each $n\geq 0$. One of or results will show that equipped with this
omega operation, each $^*$-continuous Kleene $\omega$-algebra gives rise to an 
iteration semiring-semimodule pair. This extends the known result that each
$^*$-continuous Kleene algebra is an iteration semiring. 

Part of this work is motivated by an application to so-called
\emph{energy problems} for hybrid systems.  When modeling systems with
constraints on energy consumption, it is of interest to know whether
certain states are reachable under the given energy constraints, or
whether the system admits an infinite run.  In the context of formal
modeling and verification, such problems were first taken up
in~\cite{Bouyeretaltimed}, which spurned a number of other papers on
similar problems.

We have shown in~\cite{Esiketalenergy} that most of these problems can
be covered by a new abstract notion of \emph{energy automaton}, which
is a finite automaton in which the transitions are labeled with
functions which map input energy to output energy.  The intuition is
that the transitions of an energy automaton are \emph{energy
  transformations}, so given an initial energy, one can ask whether a
final state is reachable when starting with this initial energy, or
whether there is a B\"uchi run given this initial energy.

We will show in the later parts of this paper that energy functions
form a finitary *-continuous Kleene $\omega$-algebra, hence that energy
problems can be solved using the techniques developed here.

\section{Semirings and semiring-semimodule pairs}
\label{sec-prelim}

Although we already mentioned semirings and semimodules in the introduction, 
in this section we briefly recall the main definitions from 
\cite{BerstelReutenauer,Golan} in order to make the paper self-contained. 

Recall from \cite{BerstelReutenauer,Golan} that a \emph{semiring} 
$S = (S,+,\cdot,0,1)$ consists of a commutative 
monoid $(S,+,0)$ and a monoid $(S,\cdot,1)$ such that the distributive
laws 
\begin{align*}
x(y+ z) & = xy + xz\\
(y+z)x &= yx + zx
\end{align*}
and the zero laws 
\begin{equation*}
0\cdot x = 0 = x \cdot 0
\end{equation*}
hold for all $x,y,z \in S$. It follows that the product operation distributes over 
all finite sums. 

An \emph{idempotent semiring} is a semiring $S$ whose sum operation is
idempotent, so that $x + x = x$ for all $x \in S$. Each idempotent
semiring $S$ is partially ordered by the relation $x \leq y$ iff $x +
y = y$, and then sum and product preserve the partial order and $0$ is
the least element. Moreover, for all $x,y \in S$, $x + y$ is the least
upper bound of the set $\{x,y\}$. Accordingly, in an idempotent
semiring $S$, we will usually denote the sum operation by $\vee$ and
$0$ by $\bot$.

In addition to semirings, we will also consider 
semiring-semimodule pairs. Suppose 
that $S = (S,+,\cdot,0,1)$ is a semiring and $V = (V,+,0)$ a 
commutative monoid. When $V$ is equipped with a left $S$-action
$S \times V \to V$, $(s,v) \mapsto sv$, satisfying
\begin{align*}
(s+s')v   &= sv + s'v\\
s(v + v') &= sv + sv'\\
(ss') v   &= s(s'v)\\
0s &= 0\\
s0 &= 0\\
1v &= v
\end{align*}
for all $s,s'\in S$ and $v \in V$, then we call $V$ a (unitary left) \emph{$S$-semimodule} and
$(S,V)$ a \emph{semiring-semimodule pair}.
Note that when $S$ is idempotent, then $V$ is necessarily idempotent, 
so that we use the notation  $(V,\vee,\bot)$.

\section{Free continuous Kleene $\omega$-algebras}

We have defined continuous Kleene $\omega$-algebras in the
introduction as idempotent semiring-semimodule pairs $(S,V)$ such that
$S = (S,\vee,\cdot,\bot,1)$ is a continuous Kleene algebra and $V=
(V,\vee,\bot)$ is a continuous $S$-semimodule.  Thus, equipped with
the natural order, $\leq$, $S$ and $V$ are complete lattices and the
product and the action preserve all suprema in either argument.
Moreover, there is an infinite product operation, satisfying the
compatibility and associativity conditions, which preserves all
suprema.

In this section, we offer descriptions of the free continuous Kleene $\omega$-algebras 
and the free continuous Kleene $\omega$-algebras satisfying the identity $1^\omega = \bot$. 

A homomorphism between continuous Kleene algebras preserves all operations.
A homomorphism is continuous if it preserves all suprema. 
We recall the following basic result.

\begin{theorem}
\label{thm-free0}
For each set $A$, the language semiring $(P(A^*), \vee, \cdot, \bot,1)$ 
where $\vee$ is set union, $\cdot$ is concatenation and the constants $\bot$
and $1$ are $\emptyset$ and $\{\epsilon\}$, respectively, is the free continuous 
Kleene algebra on $A$. 
\end{theorem} 

Thus, if $S$ is a continuous Kleene algebra and $h: A \to S$ is any function, 
then there is a unique continuous homomorphism $h^\sharp: A^* \to S$ extending $h$. 

In view of Theorem~\ref{thm-free0}, it is not surprising that the continuous Kleene $\omega$-algebras 
can be described using languages of finite and $\omega$-words. 

Suppose that $A$ is a set. Let $A^\omega$ denote the set of all $\omega$-words
over $A$, and let $A^\infty = A^* \cup A^\omega$. Let $P(A^*)$ denote the 
language semiring over $A$, and let $P(A^\infty)$
denote the semimodule of all subsets of $A^\infty$ equipped with the action 
of $P(A^*)$ defined by $XY = \{xy : x \in X,y\in Y\}$ for all $X \subseteq A^*$ 
and $Y\subseteq A^\infty$. We also define an infinite product by
$\prod_{n \geq 0} X_n = \{u_0u_1\ldots : u_n \in X_n\}$. 
Note that $P(A^\omega)$ is a subsemimodule of $P(A^\infty)$.

\begin{proposition}
For each set $A$, $(P(A^*),P(A^\infty))$ is a continuous Kleene $\omega$-algebra. 
\end{proposition}

Homomorphisms between continuous Kleene $\omega$-algebras $(S,V)$ and $(S',V')$ 
consist of two functions $h_S: S \to S'$ and $h_V: V \to V'$ that together 
preserve all operations. A homomorphism $(h_S,h_V)$ is continuous if 
$h_S$ and $h_V$ preserve all suprema. 

\begin{theorem}
\label{thm-free1}
$(P(A^*), P(A^\infty))$ is the free continuous Kleene $\omega$-algebra,
freely generated by $A$. 
\end{theorem} 

\proof 
Suppose that $(S,V)$ is any continuous Kleene $\omega$-algebra
an let $h: A \to S$ be a mapping. We want to show that there is 
a unique extension of $h$ to a continuous homomorphism 
$(h_S^\sharp,h_V^\sharp)$ from $(P^\omega(A), P(A^\infty))$ 
to $(S,V)$.

For each $u= a_0\ldots a_{n-1}$ in $A^*$, 
define $h_S(u) = h(a_0)\cdots h(a_{n-1})$ and $h_V(u) = h(a_0)\cdots h(a_{n-1}) 1^\omega
= \prod_{k \geq 0}b_k$, where $b_k = a_k$ for all $k < n$ and $b_k = 1$ for all $k \geq n$.
When $u = a_0a_1\ldots \in A^\omega$, define $h_V(u) = \prod_{k \geq 0} h(a_k)$. 
Note that we have $h_S(uv) = h_S(u)h_S(v)$ for all $u,v\in A^*$
and $h_S(\epsilon) = 1$. Also, $h_V(uv) = h_S(u)h_V(v)$ for all $u\in A^*$ and $v \in A^\infty$. 
Thus, $h_V(XY) = h_S(X)h_V(Y)$ for all $X \subseteq A^*$ and $Y \subseteq A^\infty$. 
Moreover, for all $u_0,u_1,\ldots$ in $A^*$, if $u_i \neq \epsilon$ for infinitely many $i$,
then $h_V(u_0u_1\ldots) = \prod_{k \geq 0} h_S(u_k)$. If on the other hand, $u_k = \epsilon$ for all $k\geq n$,
then $h_V(u_0u_1\ldots) = h_S(u_0)\cdots h_S(u_{n-1})1^\omega$.
In either case, if $X_0,X_1,\ldots \subseteq A^*$, then 
$h_V(\prod_{n \geq 0} X_n) = \prod_{n \geq 0} h_S(X_n)$. 

Suppose now that $X \subseteq A^*$ and $Y \subseteq A^\infty$. We define 
$h^\sharp_S(X) = \bigvee h_S(X)$ and $h_V^\sharp (Y) = \bigvee h_V(Y)$. It is well-known 
that $h_S^\sharp$ is a continuous semiring morphism $P(A^*) \to S$. Also, $h^\sharp_V$ 
preserves arbitrary suprema, since $h_V^\sharp (\bigcup_{i \in I} Y_i) =
\bigvee h_V(\bigcup_{i \in I} Y_i) = \bigvee \bigcup_{i \in I} h_V(Y_i) =
\bigvee_{i \in I} \bigvee h_V(Y_i) = \bigvee_{i \in I} h_V^\sharp (Y_i)$. 

We prove that the action is preserved. Let $X \subseteq A^*$ and $Y \subseteq A^\infty$. 
Then $h_V^\sharp(XY) = \bigvee h_V(XY) = \bigvee h_S(X)h_V(Y) = \bigvee h_S(X) \bigvee h_V(Y)
= h_S^\sharp(X)h_V^\sharp(Y)$.

Finally, we prove that the infinite product is preserved. 
Let $X_0,X_1,\ldots \subseteq A^*$. Then 
$h^\sharp_V(\prod_{n \geq 0}X_n) = \bigvee h_V(\prod_{n \geq 0}X_n) 
= \bigvee \prod_{n \geq 0} h_S(X_n) = \prod_{n \geq 0} \bigvee h_S(X_n) 
= \prod_{n \geq 0} h_S^\sharp(X_n)$. 

It is clear that $h_S$ extends $h$, and that $(h_S,h_V)$ is unique. 
\eop

Consider now $(P(A^*),P(A^\omega))$ with infinite product defined by $\prod_{n \geq 0} X_n =
\{u_0u_1\ldots  \in A^\omega : u_n \in X_n, \ n \geq 0\}$. It is also a continuous Kleene 
$\omega$-algebra. Moreover, it satisfies  $1^\omega = \bot$.  

\begin{lemma}
 $(P(A^*),P(A^\omega))$ is a quotient of $(P(A^*),P(A^\infty))$ 
under the continuous homomorphism $(\varphi_S,\varphi_V)$
such that $\varphi_S$  is the identity on $P(A^*)$ and $\varphi_V$ 
maps $Y \subseteq A^\infty$ to $Y \cap A^\omega$. 
\end{lemma} 

\proof Suppose that $Y_i \subseteq A^\infty$ for all $i \in I$. 
It holds $\varphi_V(\bigcup_{i \in I} Y_i) = A^\omega \cap \bigcup_{i \in I} Y_i 
= \bigcup_{i \in I} (A^\omega \cap Y_i) = \bigcup_{i \in I} \varphi_V(Y_i)$. 

Let $X \subseteq A^*$ and $Y \subseteq A^\infty$. Then $h_V(XY) = XY \cap A^\omega =
X(Y \cap A^\omega) = \varphi_S(X)\varphi_V(Y)$. 

Finally, let $X_0,X_1,\ldots \subseteq A^*$. Then $h_V(\prod_{n \geq 0} X_n) 
= \{u_0u_1\ldots \in A^\omega : u_n \in X_n\}  = \prod_{n \geq 0} h_S(X_n)$.
\eop  

\begin{lemma}
\label{lem-factors}
Suppose that $(S,V)$ is a continuous Kleene $\omega$-algebra satisfying $1^\omega = \bot$.
Let $(h_S,h_V)$ be a continuous homomorphism $(P(A^*),P(A^\infty)) \to (S,V)$. 
Then $(h_S,h_V)$ factors through $(\varphi_S,\varphi_V)$.
\end{lemma}

\proof Define $h'_S = h_S$ and $h'_V: P(A^\omega) \to V$
by $h'_V(Y) = h_V(Y)$, for all $Y \subseteq A^\omega$.
Then clearly $h_S = \varphi_S h'_S$. Moreover, $h_V = \varphi_V h'_V$, 
since for all $Y \subseteq A^\infty$, $Y\varphi_Vh'_V = (Y \cap A^\omega)h_V
= (Y \cap A^\omega)h_V \vee (Y \cap A^*)h_S 1^\omega
= (Y \cap A^\omega)h_V \vee ((Y \cap A^*)1^\omega)h_V
= ((Y \cap A^\omega) \cup (Y \cap A^*)1^\omega)h_V
= Yh_V$.

Since $(\varphi_S,\varphi_V)$ and $(h_S,h_V)$ are homomorphisms. 
so is $(h'_S,h'_V)$. It is clear  that $h'_V$ 
preserves all suprema. 
\eop 

\begin{corollary}
\label{cor-free1}
For each set $A$, $(P(A^*), P(A^\omega))$ is the free continuous Kleene $\omega$-algebra
satisfying $1^\omega = \bot$, freely generated by $A$. 
\end{corollary}

\proof Suppose that $(S,V)$ is a continuous Kleene $\omega$-algebra
satisfying $1^\omega = \bot$. Let $h: A \to S$. By Theorem~\ref{thm-free1}, there is a 
unique continuous homomorphism $(h_S,h_V) : (P(A^*),P(A^\infty)) 
\to (S,V)$ extending $h$. By Lemma~\ref{lem-factors}, $h_S$ and $h_V$ 
factor as $h_S = \varphi_S h'_S$ and $h_V = \varphi_V h_V'$, where 
$(h'_S,h'_V)$ is a continuous homomorphism $(P(A^*),P(A^\omega)) \to (S,V)$.
This homomorphism $(h'_S,h'_V)$ is the required extension of $h$ 
to a continuous homomorphism $(P(A^*),P(A^\omega)) \to (S,V)$.
Since the factorization is unique, so is this extension. 
\eop

\section{$^*$-Continuous Kleene $\omega$-algebras}
\label{sec-star cont}

In this section, we define \emph{$^*$-continuous Kleene $\omega$-algebras} 
and \emph{finitary $^*$-continuous Kleene $\omega$-algebras} 
as an extension of the \emph{$^*$-continuous Kleene algebras} of \cite{Kozen}.
We establish several basic properties of these structures, including
the existence of the supremum of certain subsets including those
corresponding to regular $\omega$-languages.

We define a $^*$-continuous Kleene $\omega$-algebra $(S,V)$ as a 
$^*$-continuous Kleene algebra $$(S,\vee ,\cdot, \bot,1,^*)$$ acting 
on a (necessarily idempotent) semimodule $V = (V,\vee,\bot)$
subject to the usual laws of unitary action as well as the following axiom \Ax0: 
\begin{equation*}
x y^*v = \bigvee_{n \geq 0} x y^n v,
\end{equation*}
for all $x,y \in S$ and $v \in V$.
Moreover, there is an infinite product operation mapping an $\omega$-sequence
$(x_0,x_1,\ldots)$ (or $\omega$-word $x_0x_1\ldots$) over $S$ to an element $\prod_{n \geq 0}x_n$ of $V$.
Thus, infinite product is a function $S^\omega \to V$, where $S^\omega$ 
denotes the set of all $\omega$-sequences over $S$. 

Infinite product is subject to the following axioms relating the 
infinite product to the other operations of Kleene $\omega$-algebras 
and operations on $\omega$-sequences. The first two axioms are 
the same as for continuous Kleene $\omega$-algebras. 
The last two ones are weaker forms of the complete 
additivity of the infinite product of continuous Kleene $\omega$-algebras. 
\begin{itemize}
\item[\Ax1:] For all $x_0,x_1,\ldots \in S$, 
$\prod_{n \geq 0}x_n = x_0\prod_{n \geq 0}y_n$, where $y_n = x_{n+1}$ for all $n \geq 0$. 
\item[\Ax2:] Suppose that $x_0,x_1,\ldots \in S$ and 
$0= n_0 \leq n_1\leq \cdots$ is a sequence which 
increases without a bound. Let $y_k = x_{n_k}\cdots x_{n_{k + 1}-1}$ for all $k \geq 0$. 
Then $\prod_{n \geq 0} x_n = \prod_{k \geq 0}y_k$.
\item[\Ax3:] For all $x_0,x_1,\ldots $ and $y,z$ in $S$, 
   $$\prod_{n \geq 0}x_n (y \vee z) = \bigvee_{x_n'\in \{y,z\}}\prod_{n \geq 0} x_nx'_n.$$
\item[\Ax4:] Suppose that $x,y_0,y_1,\ldots$ are in $S$. Then
$$\prod_{n \geq 0}x^*y_n = \bigvee_{k_n \geq 0}\prod_{n \geq 0} x^{k_n}y_n.$$
\end{itemize}
It is clear that every continuous Kleene $\omega$-algebra is $^*$-continuous. 

Some of our results will also hold for weaker structures.  We define a
\emph{finitary} $^*$-continuous Kleene $\omega$-algebra as a structure
$(S,V)$ as above, equipped with a star operation and an infinite
product $\prod_{n \geq 0}x_n$ restricted to \emph{finitary
  $\omega$-sequences} (or finitary $\omega$-words) over $S$, i.e., to
sequences $x_0,x_1,\ldots$ such that there is a finite subset $F$ of
$S$ such that each $x_n$ is a finite product of elements of $F$. (Note
that $F$ is not fixed and may depend on the sequence $x_0,
x_1,\dotsc$)  It is required that \Ax1, \Ax2 and \Ax3 hold whenever
the sequence $x_0,x_1,\ldots$ is finitary, and that \Ax4 holds
whenever the sequence $y_0,y_1,\ldots$ is finitary.

Finally, a \emph{generalized $^*$-continuous Kleene algebra} $(S,V)$
is defined as a $^*$-continuous Kleene $\omega$-algebra, but without
the infinite product (and without \Ax1--\Ax4). However, it is assumed
that \Ax0 holds.

The above axioms have a number of consequences. For example, 
if $x_0,x_1,\ldots \in S$ and $x_i = \bot$ for some $i$, then 
$\prod_{n \geq 0}x_n = \bot$.
Indeed, if $x_i = \bot$,  then $\prod_{n \geq 0}x_n = 
x_0\cdots x_i \prod_{n \geq i+1}x_n = 
\bot\prod_{n \geq i+1}x_n = \bot$. 
By \Ax1 and \Ax2, each $^*$-continuous Kleene $\omega$-algebra is 
an $\omega$-semigroup, and equipped with the natural order relation,
an ordered $\omega$-semigroup, cf. \cite[pp.~92f]{PerrinPin}.

Similarly, when $(S,V)$ is a finitary $^*$-continuous Kleene algebra 
and $x_0,x_1,\ldots$ is a finitary sequence over $S$, and if some 
$x_i$ is $\bot$, then $\prod_{n \geq 0} x_n = \bot$.

Suppose that $(S,V)$ is a $^*$-continuous Kleene $\omega$-algebra.  To
each word $w \in S^*$ there is a corresponding element $\overline{w}$
of $S$ which is the product of the letters of $w$ in the semiring $S$.
Similarly, when $w \in S^*V$, there is an element $\overline{w}$ of
$V$ corresponding to $w$.  Suppose now that a language $X \subseteq
S^*$ or $X\subseteq S^*V$.  Then we can associate with $X$ the set
$\overline X=\{\overline{w} : w\in X\}$, which is a subset of $S$ or
$V$.  The reader will probably be relieved that below we will denote
$\overline{w}$ and $\overline{X}$ by just $w$ and $X$, respectively,

The following 
facts are well-known (and follow from the result that the semirings of regular languages 
are the free $^*$-continuous Kleene algebras \cite{Kozen} (essentially shown also in \cite{Conway,Salomaa}),
and the free Kleene algebras \cite{Kozen2}).

\begin{lemma}
\label{lem-KA1}
Suppose that $S$ is a $^*$-continuous Kleene algebra. 
If $R \subseteq S^*$ is regular, then $\bigvee R$ exists. Moreover,
for all $x,y \in S$, $x (\bigvee R) y = \bigvee xRy$.
\end{lemma}

\begin{lemma}
\label{lem-KA2}
Let $S$ be a $^*$-continuous Kleene-algebra.
Suppose that $R,R_1$ and $R_2$ are regular subsets of $S^*$.
Then 
\begin{align*}
\bigvee (R_1\cup R_2) &= \bigvee R_1 \vee \bigvee R_2\\
\bigvee (R_1R_2) &= (\bigvee R_1)(\bigvee R_2)\\
\bigvee (R^*) &= (\bigvee R)^*.
\end{align*} 
\end{lemma}

In a similar way, we can prove:

\begin{lemma}
\label{lem-KOA1}
Let $(S,V)$ be a generalized $^*$-continuous Kleene-algebra.
If $R \subseteq S^*$ is regular, $x \in S$ and $v \in V$, then $x(\bigvee R)v = \bigvee xRv$.
\end{lemma}

\proof
Suppose that $R = \emptyset$. Then $x(\bigvee R)v = \bot = \bigvee xRv$.
If $R$ is a singleton set $\{y\}$, then $x(\bigvee R)v = xyv = \bigvee xRv$.
Suppose now that $R = R_1\cup R_2$ or $R = R_1R_2$,  where $R_1,R_2$ are 
regular, and suppose that our claim holds for $R_1$ 
and $R_2$. Then, if $R = R_1 \cup R_2$, 
\begin{align*}
x(\bigvee R)v &= x(\bigvee R_1 \vee \bigvee R_2)v \quad ({\rm by}\ {\rm Lemma~\ref{lem-KA2}}) \\
&= x(\bigvee R_1)v \vee x(\bigvee R_2)v \\
&= \bigvee xR_1v \vee \bigvee xR_2v\\
&= \bigvee x(R_1\cup R_2)v\\
&= \bigvee xRv,
\end{align*}
where the third equality uses the induction hypothesis. 
If $R = R_1R_2$, then 
\begin{align*}
x(\bigvee R)v &= x(\bigvee R_1)(\bigvee R_2)v \quad ({\rm by}\ {\rm Lemma~\ref{lem-KA2}})\\
&= \bigvee (xR_1 (\bigvee R_2)v)\\
&= \bigvee \{ y(\bigvee R_2)v : y \in xR_1\}\\
&=  \bigvee \{ \bigvee yR_2v : y \in xR_1\}\\
&= \bigvee xR_1R_2v\\
&= \bigvee xRv,
\end{align*} 
where the second equality uses the induction hypothesis for $R_1$ and the 
fourth the one for $R_2$.
Suppose last that $R = R_0^*$, where $R_0$ is regular and our claim holds 
for $R_0$. Then, using the previous case, it follows by induction
that 
\begin{align}
\label{eq-power}
x (\bigvee R_0^n) v&= \bigvee xR_0^nv
\end{align}
for all $n \geq 0$. Using this and \Ax0, it follows now that 
\begin{align*}
x (\bigvee R)v 
&= x (\bigvee R_0^*)y \\
&= x (\bigvee_{n \geq 0}\bigvee R_0^n)v\\
&= x (\bigvee_{n \geq 0} (\bigvee R_0)^n )v \quad ({\rm by}\ {\rm Lemma}~\ref{lem-KA2})\\
&= \bigvee_{n \geq 0} x (\bigvee R_0)^n v \quad ({\rm by} {\rm \Ax0})\\
&= \bigvee_{n \geq 0} x (\bigvee R_0^n)v \quad ({\rm by}\ {\rm Lemma}~\ref{lem-KA2})\\
&= \bigvee_{n \geq 0} \bigvee xR_0^n v \quad  ({\rm by} \ (\ref{eq-power}))\\
&= \bigvee xR_0^*v\\
&= \bigvee xRv.
\end{align*}
The proof is complete.
\eop

We can also prove:

\begin{lemma}
\label{lem-KOA2}
Let $(S,V)$ be a $^*$-continuous Kleene $\omega$-algebra.
Suppose that the languages $R_0,R_1, \ldots \subseteq S^*$ are regular 
and form a finite set $\R$ of languages. Moreover, let $x_0,x_1,\ldots \in S$. 
Then
$$\prod_{n \geq 0} x_n(\bigvee R_n) =  \bigvee \prod_{n \geq 0} x_nR_n .
$$
\end{lemma}

\proof
If one of the $R_i$ is empty, our claim is clear since both sides are equal to
$\bot$, so we suppose they are all nonempty.

Below we will suppose that each regular language comes with a fixed decomposition 
having a minimal number of operations needed to obtain the language from the empty 
set and singleton sets. 
For a regular language $R$, let $|R|$ denote the minimum number of operations
needed to construct it. When $\R$ is a finite set of regular languages, let $\R'$ 
denote the set of non-singleton languages in it. Let 
$|\R| = \sum_{R \in \R'} 2^{|R|}$. Our definition ensures that if 
$\R = \{R,R_1,\ldots,R_n\}$ and $R = R' \cup R''$ or $R = R'R''$ 
according to the fixed minimal decomposition of $R$, and if $\R' = 
\{R',R'',R_1,\ldots,R_n\}$, then $|\R'| < |\R|$. Similarly, if $R = R_0^*$
by the fixed minimal decomposition
and $\R'=\{R_0,R_1,\ldots,R_n\}$, then $|\R'| < |\R|$.

We will argue by induction on $|\R|$.

When $|\R| = 0$, then $\R$ consists of singleton languages and our claim 
follows from \Ax3. 
Suppose that $|\R| > 0$. Let $R$ be a non-singleton language appearing in $\R$. 
If $R$ appears only a finite number of times among the $R_n$, then there is some 
$m$ such that $R_n$ is different from $R$ for all $n \geq m$. Then,  
\begin{align*}
\prod_{n \geq 0} x_n(\bigvee R_n) &= 
\prod_{i < m} x_i(\bigvee R_i)\prod_{n \geq m}x_n(\bigvee R_n)\quad ({\rm by}\ {\rm \Ax1})\\
&= 
(\bigvee x_0R_0\cdots x_{n-1}R_{n-1})  \prod_{n \geq m}x_n(\bigvee R_n)\quad ({\rm by}\ 
{\rm Lemma~\ref{lem-KA2}})\\
&=
\bigvee (x_0R_0\cdots x_{n-1}R_{n-1} \prod_{n \geq m}x_n(\bigvee R_n))\quad 
({\rm by}\ {\rm Lemma~\ref{lem-KOA1}}) \\
&= 
\bigvee \{ y  \prod_{n \geq m}x_n(\bigvee R_n) : y \in x_0R_0\cdots x_{n-1}R_{n-1}\}\\
&= 
\bigvee \{\bigvee y \prod_{n \geq m} x_nR_n  : y \in x_0R_0\cdots x_{n-1}R_{n-1}\}\\
&= 
\bigvee \prod_{n \geq 0} x_nR_n,
\end{align*}
where the passage from the 4th line to the 5th 
uses induction hypothesis and \Ax1.

Suppose now that $R$ appears an infinite number of times among the $R_n$. 
Let $R_{i_1},R_{i_2},\ldots$ be all the occurrences of 
$R$ among the $R_n$. Define 
\begin{align*}
y_0 &=  x_0 (\bigvee R_0)\cdots (\bigvee R_{i_1-1})x_{i_1}\\
y_j &= x_{i_j + 1}(\bigvee R_{i_j +  1}) \cdots (\bigvee R_{i_{j + 1}-1})x_{i_{j+ 1}},
\end{align*}
for $j \geq 1$.  Similarly, define
\begin{align*}
Y_0 &= x_0 R_0 \cdots  R_{i_1-1} x_{i_1}\\ 
Y_j &=  x_{i_j +  1} R_{i_j + 1} \cdots R_{i_{j + 1}-1}x_{i_{j + 1}},
\end{align*}
for all $j \geq 1$.
It follows from Lemma~\ref{lem-KA2} that 
\begin{equation*}
y_j = \bigvee Y_j
\end{equation*}
for all $j \geq 0$.
Then  
\begin{equation}
\label{eq-YjR}
\prod_{n \geq 0} x_n(\bigvee R_n)
= 
\prod_{j \geq 0} y_j(\bigvee R),
\end{equation} 
by \Ax2, and 
\begin{equation*}
\prod_{n \geq 0} x_n R_n
= 
\prod_{j \geq 0} Y_j R.
\end{equation*}

If $R = R' \cup R''$, 
then:
\begin{align*}
\prod_{n \geq 0} x_n(\bigvee R_n) 
&=
\prod_{j \geq 0} y_j(\bigvee (R' \cup R''))\quad 
({\rm by}\ \eqref{eq-YjR})\\
&= 
\prod_{j \geq 0} y_j(\bigvee R' \vee  \bigvee R'')\quad ({\rm by} \ {\rm Lemma~\ref{lem-KA2}})\\
&=
\bigvee_{z_j \in \{\bigvee R', \bigvee R''\}} \prod_{j \geq 0} y_jz_j\quad ({\rm by}\ {\rm \Ax3})\\
&= 
\bigvee_{z_j \in \{\bigvee R',\bigvee R''\}} \bigvee \prod_{j \geq 0} Y_jz_j\\
&= 
\bigvee_{Z_j \in \{R',R''\}} \bigvee \prod_{j \geq 0} Y_jZ_j\\
&= 
\bigvee \prod_{n \geq 0} x_n(R'\cup R'')\\
&= 
\bigvee \prod_{n\geq 0}x_nR,
\end{align*}
where the 4th and 5th equalities hold by the induction hypothesis and \Ax2.

Suppose now that $R = R'R''$. Then applying the induction hypothesis 
almost directly we have
\begin{align*}
\prod_{n \geq 0} x_n(\bigvee R_n)
&= 
\prod_{j \geq 0} y_j(\bigvee R'R'')\\
&= 
\prod_{j \geq 0} y_j(\bigvee R')(\bigvee R'') \quad ({\rm by} \ {\rm Lemma~\ref{lem-KA2}})\\
&=
\bigvee \prod_{j \geq 0} Y_j(\bigvee R')(\bigvee R'')\\
&=
\bigvee \prod_{j \geq 0} Y_jR'R''\\
&= 
\bigvee \prod_{n \geq 0} x_nR'R''\\
&= 
\bigvee \prod_{n\geq 0}x_nR,
\end{align*}
where the third and fourth equalities come from the induction hypothesis and \Ax2.

The last case to consider is when $R= T^*$, where $T$ is 
regular. We argue as follows:
\begin{align*}
\prod_{n \geq 0} x_n(\bigvee R_n)
&= 
\prod_{j \geq 0} y_j(\bigvee T^*)\\
&= 
\prod_{j \geq 0}y_j(\bigvee T)^*\quad ({\rm by} \ {\rm Lemma~\ref{lem-KA2}})\\
&=
\bigvee_{k_0,k_1,\ldots}\prod_{j \geq 0} y_j(\bigvee T)^{k_j}\quad ({\rm by}\ {\rm \Ax1}\ {\rm and}\ {\rm \Ax4})\\
&=
\bigvee_{k_0,k_1,\ldots}\bigvee \prod_{j \geq 0} Y_j(\bigvee T)^{k_j}
\\
&=
\bigvee_{k_0,k_1,\ldots}\bigvee \prod_{j \geq 0} Y_j T^{k_j}
\\
&=\bigvee_{j \geq 0}Y_jT^*\\
&= 
\bigvee_{j \geq 0} Y_jR_j\\
&= 
\bigvee_{n \geq 0} x_nR_n,
\end{align*}
where the 4th and 5th equalities follow from the induction hypothesis and \Ax2.
The proof is complete.
\eop

By the same proof, we have the following version of Lemma~\ref{lem-KOA2} for the finitary case:

\begin{lemma}
\label{lem-KOA2weak}
Let $(S,V)$ be a finitary $^*$-continuous Kleene $\omega$-algebra.
Suppose that the languages $R_0,R_1, \cdots \subseteq S^*$ are regular 
and form a finite set $\R$ of languages. Moreover, let
$x_0,x_1,\ldots$ is a finitary sequence of elements of  $S$. 
Then
$$\prod_{n \geq 0} x_n(\bigvee R_n) =  \bigvee \prod_{n \geq 0} x_nR_n. 
$$
\end{lemma}

Note that each sequence $x_0,y_0,x_1,y_1,\ldots $ with $y_n \in R_n$
is finitary. 

As an immediate consequence of Lemma~\ref{lem-KOA2weak}, we get the following result:

\begin{corollary}
\label{cor-infprodreg} 
Let $(S,V)$ be a finitary $^*$-continuous Kleene $\omega$-algebra.
Suppose that $R_0,R_1,\ldots \subseteq S^*$ are regular and form a 
finite set of languages. Then $\bigvee \prod_{n \geq 0} R_n$ 
exists and is equal to  $\prod_{n \geq 0} \bigvee R_n$. 
\end{corollary}

Suppose that $(S,V)$ is a $^*$-continuous Kleene $\omega$-algebra. 
When $v = x_0x_1\ldots \in S^\omega$ is an $\omega$-word over $S$, 
it naturally determines the element $\prod_{n \geq 0}x_n$ of $V$. 
Thus, any subset $X$ of $S^\omega$ determines a subset of $V$.
Using this convention, Lemma~\ref{lem-KOA2} may be rephrased as follows. For any 
$^*$-continuous Kleene $\omega$-algebra $(S,V)$, $x_0,x_1,\ldots \in S$ and 
regular sets $R_0,R_1,\ldots \subseteq S^*$ that form a finite set, 
it holds that 
$\prod_{n \geq 0} x_n(\bigvee R_n) =  \bigvee  X$ where $X \subseteq S^\omega$ is the 
set of all $\omega$-words $x_0y_0x_1y_1\ldots $ with $y_i \in R_i$ for all $i \geq 0$,
i.e., $X = x_0R_0x_1R_1\ldots$ Similarly, Corollary~\ref{cor-infprodreg} asserts 
that if a subset of $V$ corresponds to an infinite product over a finite collection
of ordinary regular languages in $S^*$, then the supremum of this set exists.

In any (finitary or non-finitary) $^*$-continuous Kleene $\omega$-algebra 
$(S,V)$, 
we define an \emph{$\omega$-power operation}
$S \to V$ by $x^\omega = \prod_{n \geq 0}x$ for all $x \in S$. 
From the axioms we immediately have:

\begin{corollary}
Suppose that $(S,V)$ is a $^*$-continuous Kleene $\omega$-algebra
or a finitary $^*$-continuous Kleene $\omega$-algebra.
Then the following hold for all $x,y \in S$:
\begin{align*}
x^\omega &= xx^\omega\\
(xy)^\omega &= x(yx)^\omega\\
x^\omega &= (x^n)^\omega,\quad n \geq 2.
\end{align*}
\end{corollary}

Thus, each $^*$-continuous Kleene $\omega$-algebra gives rise to a Wilke algebra,
even an ordered Wilke algebra \cite{PerrinPin,Wilke}.

\begin{lemma}
\label{lem-KOA3}
Let $(S,V)$ be a $^*$-continuous Kleene $\omega$-algebra
or a finitary $^*$-continuous Kleene $\omega$-algebra. 
Suppose that $R \subseteq S^\omega$ is $\omega$-regular. Then $\bigvee R$ exists in $V$. 
\end{lemma}

\proof
It is well-known that $R$ can be written as a finite union of sets 
of the form $R_0 (R_1)^\omega$ where $R_0,R_1\subseteq S^*$ are regular, moreover,
$R_1$ does not contain the empty word. It suffices to show that $\bigvee R_0(R_1)^\omega$ 
exists. But this holds by Corollary~\ref{cor-infprodreg}. 
\eop

\begin{lemma}
\label{lem-KOA4}
Let $(S,V)$ be a $^*$-continuous Kleene $\omega$-algebra.
 For all $\omega$-regular sets $R_1,R_2\subseteq S^\omega$ and regular sets 
$R \subseteq S^*$ it holds that
\begin{align*}
\bigvee (R_1 \cup R_2) &= \bigvee R_1 \vee \bigvee R_2\\
\bigvee (RR_1) &= (\bigvee R)(\bigvee R_1).
\end{align*}
And if $R$ does not contain the empty word, then 
\begin{equation*}
\bigvee R^\omega = (\bigvee R)^\omega.
\end{equation*}
\end{lemma}

\proof
The first claim is clear. The second follows from Lemma~\ref{lem-KOA1}. For the last,
see the proof of Lemma~\ref{lem-KOA3}. 
\eop

\section{Free finitary $^*$-continuous Kleene $\omega$-algebras}

Recall that for a set $A$, $R(A^*)$ denotes the collection of all
regular languages in $A^*$.  It is well-known that $R(A^*)$, equipped
with the usual operations, is a $^*$-continuous Kleene algebra on
$A$. Actually, $R(A^*)$ is characterized up to isomorphism by the
following universal property.

Call a function $f: S \to S'$ between $^*$-continuous Kleene algebras
a $^*$-continuous homomorphism if it preserves all operations
including star, so that it preserves the suprema of subsets of $S$ of
the form $\{x^n: n\geq 0\}$, where $x \in S$.

\begin{theorem} \cite{Kozen2}
For each set $A$, $R(A^*)$ is the free $^*$-continuous Kleene algebra,
freely generated by $A$. 
\end{theorem} 

Thus, if $S$ is any $^*$-continuous Kleene algebra
and $h$ is any mapping of $A$ into $S$, then $h$ has a unique extension to
a $^*$-continuous Kleene algebra homomorphism $h^\sharp: R(A^*) \to S$. 

Now let $R'(A^\infty)$ denote the collection of all subsets of $A^\infty$
which are finite unions of sets 
of the form $\prod_{n \geq 0}R_n$, where each $R_n \subseteq A^*$ is regular, and the 
set $\{R_0,R_1,\ldots\}$ is finite. Note that $R'(A^\infty)$ contains the empty
set and is closed under finite unions. Moreover, when $Y \in R'(A^\infty)$ 
and $u = a_0a_1\ldots \in Y \cap A^\omega$, then the alphabet of $u$ 
is finite, i.e., the set $\{a_n : n \geq 0\}$ is finite.

Also, $R'(A^\infty)$ is closed under 
the action of $R(A^*)$ inherited from $(P(A^*), P(A^\infty))$.
The infinite product of a sequence of regular languages in $R(A^*)$ 
is not necessarily contained in $R'(A^\infty)$, but 
by definition it contains all  infinite products of finitary sequences 
over $R(A^*)$. 

\begin{example}
Let $A = \{a,b\}$ and consider the set $\{aba^2b\ldots a^nb\ldots\}$ 
containing a single $\omega$-word. It can be written as an infinite
product of singleton subsets of $A^*$, but it cannot be written 
as an infinite product $R_0R_1\ldots$ of regular languages in $A^*$
such that the set $\{R_0,R_1,\ldots\}$ is finite. 
\end{example}

\begin{proposition}
$(R(A^*), R'(A^\infty))$ is a finitary $^*$-continuous Kleene $\omega$-algebra. 
\end{proposition}

\proof This is clear from the fact that $(P(A^*),P(A^\infty))$ is a continuous 
Kleene $\omega$-algebra and that $R(A^*)$ is a $^*$-continuous semiring. 
\eop

\begin{theorem}
\label{thm-free2}
$(R(A^*), R'(A^\infty))$ is the free finitary $^*$-continuous Kleene $\omega$-algebra,
freely generated by $A$. 
\end{theorem} 

\proof Our proof is modeled after the proof of Theorem~\ref{thm-free1}. 

Suppose that $(S,V)$ is any finitary $^*$-continuous Kleene
$\omega$-algebra and let $h: A \to S$ be a mapping. For each $u=
a_0\ldots a_{n-1}$ in $A^*$, define $h_S(u) = h(a_0)\cdots h(a_{n-1})$
and $h_V(u) = h(a_0)\cdots h(a_{n-1}) 1^\omega = \prod_{k \geq 0}b_k$,
where $b_k = a_k$ for all $k < n$ and $b_k = 1$ for all $k \geq n$.
When $u = a_0a_1\ldots \in A^\omega$ whose alphabet is finite, define
$h_V(u) = \prod_{k \geq 0} h(a_k)$. This infinite product exists in
$R'(A^\infty)$.

Note that we have $h_S(uv) = h_S(u)h_S(v)$ for all $u,v\in A^*$, and
$h_S(\epsilon) = 1$.  And if $u \in A^*$ and $v \in A^\infty$ such
that the alphabet of $v$ is finite,
then $h_V(uv) = h_S(u)h_V(v)$.  Also, $h_V(XY) = h_S(X)h_V(Y)$ for all
$X \subseteq A^*$ in $R(A^*)$ and $Y \subseteq A^\infty$ in
$R'(A^\infty)$.

Moreover, for all $u_0,u_1,\ldots$ in $A^*$, if $u_i \neq \epsilon$ for infinitely many $i$,
such that the alphabet of $u_0u_1\ldots$ is finite, then 
$h_V(u_0u_1\ldots) = \prod_{k \geq 0} h_S(u_k)$. If on the other hand, $u_k = \epsilon$ for all $k\geq n$,
then $h_V(u_0u_1\ldots) = h_S(u_0)\cdots h_S(u_{n-1})1^\omega$.
In either case, if $X_0,X_1,\ldots \subseteq A^*$ are regular and form a finitary 
sequence,  then the sequence $h_S(X_0), h_S(X_1),\ldots $ is also finitary
as is each infinite word in $\prod_{n \geq 0} X_n$, and
$h_V(\prod_{n \geq 0} X_n) = \prod_{n \geq 0} h_S(X_n)$. 

Suppose now that $X \subseteq A^*$ is regular and $Y \subseteq A^\infty$
is in $R'(A^\infty)$.  We define 
$h^\sharp_S(X) = \bigvee h_S(X)$ and $h_V^\sharp (Y) = \bigvee h_V(Y)$. It is well-known 
that $h_S^\sharp$ is a $^*$-continuous semiring morphism $R(A^*) \to S$. Also, $h^\sharp_V$ 
preserves finite suprema, since when $I$ is finite, $h_V^\sharp (\bigcup_{i \in I} Y_i) =
\bigvee h_V(\bigcup_{i \in I} Y_i) = \bigvee \bigcup_{i \in I} h_V(Y_i) =
\bigvee_{i \in I} \bigvee h_V(Y_i) = \bigvee_{i \in I} h_V^\sharp (Y_i)$. 

We prove that the action is preserved. Let $X \in R(A^*)$ and $Y \in R'(A^\infty)$. 
Then $h_V^\sharp(XY) = \bigvee h_V(XY) = \bigvee h_S(X)h_V(Y) = \bigvee h_S(X) \bigvee h_V(Y)
= h_S^\sharp(X)h_V^\sharp(Y)$.

Finally, we prove that infinite product of finitary sequences is preserved. 
Let $X_0,X_1,\ldots $ be a finitary sequence of regular languages in $R(A^*)$.  Then,
using Corollary~\ref{cor-infprodreg}, 
$h^\sharp_V(\prod_{n \geq 0}X_n) = \bigvee h_V(\prod_{n \geq 0}X_n) 
= \bigvee \prod_{n \geq 0} h_S(X_n) = \prod_{n \geq 0} \bigvee h_S(X_n) 
= \prod_{n \geq 0} h_S^\sharp(X_n)$. 

It is clear that $h_S$ extends $h$, and that $(h_S,h_V)$ is unique. 
\eop

Consider now $(R(A^*),R'(A^\omega))$ equipped with the infinite product operation 
$\prod_{n \geq 0} X_n =
\{u_0u_1 \in A^\omega : u_n \in X_n, \ n \geq 0\}$, defined on finitary sequences $X_0,X_1,\ldots$
of languages in $R'(A^*)$. 

\begin{proposition}
$(R(A^*),R'(A^\omega))$
is a $^*$continuous Kleene $\omega$-algebra. Moreover, it satisfies  $1^\omega = \bot$.  
\end{proposition} 

\begin{lemma}
$(R(A^*),R'(A^\omega))$ is a quotient of $(R(A^*),R'(A^\infty))$ 
under the $^*$-continuous homomorphism $(\varphi_S,\varphi_V)$
such that $\varphi_S$ is the identity function on $P(A^*)$ and $\varphi_V$ 
maps $Y \in R'(A^\omega)$ to $Y \cap A^\omega$.
\end{lemma} 

\begin{lemma}
\label{lem-factor2}
Suppose that $(S,V)$ is a finitary $^*$-continuous Kleene $\omega$-algebra satisfying $1^\omega = \bot$.
Let $(h_S,h_V)$ be a $^*$-continuous homomorphism $(R(A^*),R'(A^\infty)) \to (S,V)$. 
Then $(h_S,h_V)$ factors through $(\varphi_S,\varphi_V)$.
\end{lemma}

\begin{corollary}
\label{cor-free2}
For each set $A$, $(R(A^*), R'(A^\omega))$ is the free finitary $^*$-continuous Kleene $\omega$-algebra
satisfying $1^\omega = \bot$, freely generated by $A$. 
\end{corollary} 

\proof This follows from Theorem~\ref{thm-free2} using Lemma~\ref{lem-factor2}. \eop

\section{$^*$-continuous Kleene $\omega$-algebras and iteration 
semiring-semimodule pairs}

In this section, our aim is to relate $^*$-continuous Kleene
$\omega$-algebras to iteration semiring-semimodule pairs. Our main
result will show that every (finitary or non-finitary) $^*$-continuous
Kleene $\omega$-algebra is an iteration semiring-semimodule pair.

Some definitions are in order. 
Suppose that $S$ is a semiring. Following \cite{BEbook}, 
we  call $S$ a \emph{Conway semiring} if $S$ 
is equipped with a star operation $^*: S \to S$ satisfying 
\begin{align*}
(x+y)^* &= (x^*y)^*x^*\\
(xy)^* &= 1 + x(yx)^*y
\end{align*} 
for all $x,y\in S$. 

It is known that if $S$ is a Conway semiring, then for each $n \geq 1$,
so is the semiring $S^{n \times n}$ of all $n \times n$-matrices over $S$ 
with the usual sum and product operations and the star operation defined by induction 
on $n$ so that if $n > 1$ and 
$M = \left(\begin{smallmatrix} a & b \\
c & d  \end{smallmatrix}\right)$, where $a$ and $d$ are square matrices of dimension $< n$,
then 
\begin{equation*}
M^* = 
\begin{pmatrix} (a +  bd^*c)^* & (a +  bd^*c)^*bd^*\\
(d +  ca^*b)^*ca^* & (d +  ca^*b)^*  \end{pmatrix}.
\end{equation*} 
It is known that the above inductive definition does not depend on
how $M$ is split into 4 submatrices.

Suppose that $S$ is a Conway semiring and $G = \{g_1,\ldots,g_n\}$ is a finite group
of order $n$. For each $x_{g_1},\ldots,x_{g_n}\in S$, consider the $n\times n$ 
matrix $M_G = M_G(x_{g_1},\ldots,x_{g_n})$ whose $i$th row is $(x_{g_i^{-1}g_1},\ldots,x_{g_i^{-1}g_n})$,
for $i = 1,\ldots,n$,  
so that each row (and column) is a permutation of the first row.
We say that the group identity \cite{Conway} associated with $G$ holds in 
$S$ if for each $x_{g_1},\ldots,x_{g_n}$, the first (and then any) row sum 
of $M_G^*$ is $(x_{g_1}+ \cdots + x_{g_n})^*$. Finally, we call $S$ 
an \emph{iteration semiring} \cite{BEbook,EsItsem} if all group identities hold in $S$.

Classes of examples of (idempotent) iteration semirings are given by
the continuous Kleene algebras and the $^*$-continuous Kleene algebras
defined in the introduction.  As mentioned above, the language
semirings $P(A^*)$ and the semirings $P(A \times A)$ of binary
relations are continuous and hence also $^*$-continuous Kleene
algebras, and the semirings $R(A^*)$ of regular languages are
$^*$-continuous Kleene algebras.

When $S$ is a $^*$-continuous Kleene algebra and $n$ is a nonnegative
integer, then the matrix semiring $S^{n \times n}$ is also a
$^*$-continuous Kleene algebra and hence an iteration semiring,
cf.\cite{Kozen}.  The star operation is defined by
\begin{equation*}
M^*_{i,j}  = \bigvee_{m \geq 0,\ 1 \leq k_1,\ldots,k_m \leq n} M_{i,k_1}M_{k_1,k_2}\cdots M_{k_m,j},
\end{equation*}
for all $M \in S^{n \times n}$ and $1 \leq i,j \leq n$.  It is not
immediately obvious to prove that the above supremum exists. The fact
that $M^*$ is well-defined can be established by induction on $n$
together with the well-known matrix star formula mentioned above:
Suppose that $n \geq 2$ and $M = \left(\begin{smallmatrix} a & b \\
  c & d \end{smallmatrix}\right)$, where $a$ and $d$ are square matrices of
dimension $< n$, then
\begin{equation*}
M^* =
\begin{pmatrix} (a \vee  bd^*c)^* & (a \vee  bd^*c)^*bd^*\\
(d \vee  ca^*b)^*ca^* & (d \vee  ca^*b)^*  \end{pmatrix}.
\end{equation*}

A semiring-semimodule pair $(S,V)$ is a \emph{Conway semiring-semimodule pair}
if it is equipped with a star operation $^*: S \to S$ and 
an omega operation $^\omega: S \to V$ such that $S$ is a 
Conway semiring and the following hold for all $x,y\in S$:
\begin{align*}
(x + y)^\omega &= (x^*y)^*x^\omega + (x^*y)^\omega\\
(xy)^\omega &= x(yx)^\omega.
\end{align*}
It is known that when $(S,V)$ is a Conway semiring-semimodule pair,
then so is $(S^{n\times n},V^n)$ for each $n$, where $V^n$ denotes the
$S^{n \times n}$-semimodule of all $n$-dimensional (column) vectors
over $V$ with the action of $S^{n \times n}$ defined similarly to
matrix-vector product, and where the omega operation is defined by
induction so that when
$n > 1$ and $M$ is the matrix $\left(\begin{smallmatrix} a & b \\
    c & d \end{smallmatrix}\right)$, where $a$ and $d$ are square
matrices of dimension $< n$, then
\begin{equation*}
  M^\omega = 
  \begin{pmatrix}
    (a +  bd^*c)^\omega \,+\, (a +  bd^*c)^*bd^\omega\\
    (d +  ca^*b)^\omega \,+\, (d +  ca^*b)^*ca^\omega
  \end{pmatrix} .
\end{equation*}

We also define \emph{iteration semiring-semimodule pairs}
\cite{BEbook,EsikKuich} as those Conway semi\-ring-semimodule pairs
such that $S$ is an iteration semiring and the omega operation
satisfies the following condition: let $M_G =
M_G(x_{g_1},\ldots,x_{g_n})$ with $x_{g_1},\ldots,x_{g_n}\in S$ for a
finite group $G = \{g_1,\ldots,g_n\}$ of order $n$, then the first
(and hence any) entry of $M_G^\omega$ is equal to $(x_{g_1}+\cdots +
x_{g_n})^\omega$.

Examples of (idempotent) iteration semiring-semimodule pairs include
the semiring-semimodule pairs $(P(A^*), P(A^\omega))$ of languages and
$\omega$-languages over an alphabet $A$, mentioned in the
introduction. The omega operation is defined by $X^\omega = \prod_{n
  \geq 0}X_n$, where $X_n = X$ for all $n \geq 0$.  More generally, it
is known that every continuous Kleene $\omega$-algebra gives rise to
an iteration semiring-semimodule pair. The omega operation is defined
as for languages: $x^\omega = \prod_{n \geq 0}x_n$ with $x_n = x$ for
all $n \geq 0$.

Other not necessarily idempotent examples include the \emph{complete} 
and the \emph{(symmetric) bi-inductive semiring-semimodule pairs} of 
\cite{EsikKuich1and2,EsikKuich}.

Suppose now that $(S,V)$ is a $^*$-continuous Kleene $\omega$-algebra. 
Then for each $n\geq 1$, $(S^{n \times n}, V^n)$ is a semiring-semimodule 
pair. The action of $S^{n \times n}$ on $V^n$ is defined by 
a formula similar to matrix multiplication (viewing the 
elements of $V^n$ as column matrices). It is easy to see
that $(S^{n \times n}, V^n)$ is a generalized $^*$-continuous Kleene algebra
for each $n \geq 1$.

Suppose that $n \geq 2$. We would like to define an infinite product 
operation $(S^{n \times n})^\omega \to V^n$ on matrices 
in $S^{n \times n}$ by 
\begin{equation*}
(\prod_{m \geq 0}M_m)_i = \bigvee_{1 \leq i_1,i_2,\ldots \leq n}(M_0)_{i,i_1}(M_1)_{i_1,i_2}\cdots 
\end{equation*}
for all $1 \leq i \leq n$.
However, unlike in the case of complete semiring-semimodule pairs 
\cite{EsikKuich}, the supremum on the right-hand side may not exist. Nevertheless it is possible to define an 
omega operation $S^{n \times n} \to V^n$ 
and to turn $(S^{n \times n},V^n)$ into an iteration semiring-semimodule pair.

\begin{lemma}
\label{lem-MAT1}
Let $(S,V)$ be a (finitary) $^*$-continuous Kleene $\omega$-algebra.
Suppose that $M \in S^{n \times n}$, where $n \geq 2$.
Then for every $1\leq i \leq n$,
\begin{equation*}
(\prod_{m \geq 0}M)_i = \bigvee_{1 \leq i_1,i_2,\ldots \leq n}M_{i,i_1}M_{i_1,i_2}\cdots 
\end{equation*}
exists, so that we \emph{define} $M^\omega$ by the above equality. 

Moreover, when $M = \left(\begin{smallmatrix} a & b \\
c & d  \end{smallmatrix}\right)$, where $a$ and $d$ are square matrices
of dimension $< n$, then 
\begin{equation}
  \label{eq-matomega}
  M^\omega = 
  \begin{pmatrix}
    (a \vee bd^*c)^\omega \,\vee\,  (a \vee  bd^*c)^*bd^\omega\\
    (d \vee  ca^*b)^\omega \,\vee\,  (d \vee  ca^*b)^*ca^\omega
  \end{pmatrix}.
\end{equation} 
\end{lemma}

\proof
Suppose that $n = 2$. Then by Corollary~\ref{cor-infprodreg},
$(a \vee bd^*c)^\omega$ is the supremum 
of the set of all infinite products $A_{1,i_1}A_{i_1,i_2}\cdots$  containing 
$a$ or $c$ infinitely often, and $(a \vee  bd^*c)^*bd^\omega$ is the supremum 
of the set of all infinite products $A_{1,i_1}A_{i_1,i_2}\cdots$  containing 
$a$ and $c$ only finitely often. 
Thus, $(a \vee bd^*c)^\omega \vee  (a \vee  bd^*c)^*bd^\omega$
is the supremum of the set of all infinite products $A_{1,i_1}A_{i_1,i_2}\cdots$.
Similarly, $(d \vee  ca^*b)^\omega  \vee  (d \vee  ca^*b)^*ca^\omega$
is the supremum of the set of all infinite products $A_{2,i_1}A_{i_1,i_2}\cdots$.

The proof of the induction step is similar. Suppose that $n > 2$, and let $a$ be $k \times k$.
Then by induction hypothesis, for every $i$ with $1 \leq i \leq k$, the $i$th 
component of $(a \vee  bd^*c)^\omega$ is the supremum of the set of all infinite products 
$A_{i,i_1}A_{i_1,i_2}\cdots$ containing an entry of $a$ or $c$ infinitely often,
whereas the $i$th component of $(a \vee  bd^*c)^*bd^\omega$ is the supremum of all infinite products 
$A_{i,i_1}A_{i_1,i_2}\cdots$ containing entries of $a$ and $c$ only finitely often. 
Thus, the $i$th component of $(a \vee bd^*c)^\omega \vee  (a \vee  bd^*c)^*bd^\omega$
is the supremum of the set of all infinite products $A_{i,i_1}A_{i_1,i_2}\cdots$.
A similar fact holds for $(d \vee  ca^*b)^\omega  \vee  (d \vee  ca^*b)^*ca^\omega$.
The proof is complete.
\eop

\begin{theorem}
Every  (finitary) $^*$-continuous Kleene $\omega$-algebra is an iteration semi\-ring-semi\-module pair.
\end{theorem}

\proof
Suppose that $(S,V)$ is a finitary $^*$-continuous Kleene $\omega$-algebra.
Then 
\begin{equation*}
(x \vee y)^\omega = (x^*y)^\omega \vee (x^*y)^*x^\omega, 
\end{equation*}
since by Lemma~\ref{lem-KA2} and Lemma~\ref{lem-KOA4}, $(x^*y)^\omega$ is the supremum of the set of all infinite products 
over $\{x,y\}$ containing $y$ infinitely often, and  $(x^*y)^*x^\omega$
is the supremum of the set of infinite products over $\{x,y\}$ containing $y$ 
finitely often. Thus, $(x^*y)^\omega \vee (x^*y)^*x^\omega$ is equal to 
$(x \vee y)^\omega$, which by \Ax3 is the supremum of all infinite products over $\{x,y\}$.
As noted above, also 
\begin{equation*}
(xy)^\omega = x(yx)^\omega
\end{equation*}
for all $x,y \in S$.
Thus, $(S,V)$ is a Conway semiring-semimodule pair and hence so is 
each $(S^{n \times n}, V^n)$.

To complete the proof of the fact that $(S,V)$ is an iteration 
semiring-semimodule pair, suppose that $x_1,\ldots,x_n \in S$,
and let $x = x_1 \vee \cdots \vee x_n$. Let $A$ be an $n \times n$ 
matrix whose rows are permutations of the $x_1,\ldots,x_n$. 
We need to prove that each component of $A^\omega$ 
is $x^\omega$. We use Lemma~\ref{lem-MAT1} and \Ax3
to show that both are equal to the supremum of the set of
all infinite products over the set $X = \{x_1,\ldots,x_n\}$.

By Lemma~\ref{lem-MAT1}, for each $i_0 = 1,\ldots,n$, the $i_0$th row of $A^\omega$ is
$\bigvee_{i_1,i_2,\ldots} a_{i_0,i_1}a_{i_1,i_2}\cdots$. 
It is clear that each infinite product $a_{i_0,i_1}a_{i_1,i_2}\cdots$
is an infinite product over $X$. Suppose now that 
$x_{j_0}x_{j_1}\cdots$ is an infinite product over $X$. 
We define by induction on $k \geq 0$ an index $i_{k+1}$ 
such that $a_{i_k,i_{k+1}} = x_{j_k}$.
Suppose that $k = 0$. Then let $i_1$ be such that 
$a_{i_0,i_1} = x_{j_0}$.  Since $x_{j_0}$ 
appears in the $i_0$th row, there is such 
an $i_1$. Suppose that $k > 0$ and that $i_k$ has already been defined. 
Since $x_{j_k}$ appears in the $i_k$th 
row, there is some $i_{k+1}$ with $a_{i_k,i_{k+1}} = x_{j_k}$. 
We have completed the proof of the fact that the $i_0$th entry of $A^\omega$ is the supremum of the set of
all infinite products over the set $X = \{x_1,\ldots,x_n\}$.

Consider now $x^\omega = xx\cdots$. We use induction on $n$ 
to prove that $x^\omega$  is also the supremum of the set of
all infinite products over the set $X = \{x_1,\ldots,x_n\}$.
When $n = 1$ this is clear. Suppose now that $n > 0$ 
and that the claim is true for $n-1$. Let $y = x_1\vee \cdots \vee x_{n-1}$ so that $x = y \vee x_n$. We have:
\begin{align*}
x^\omega &= (y \vee x_n)^\omega\\
&= (x_n^* y)^* x_n^\omega  \vee (x_n^*y)^\omega\\
&= (x_n^* y)^* x_n^\omega  \vee (x_n^*x_1 \vee \cdots \vee x_n^*x_{n-1})^\omega.
\end{align*}
Now 
\begin{equation*}
(x_n^* y)^* x_n^\omega =
\bigvee_{k,m_1,\ldots,m_k\geq 0} x_n^{m_1}y\cdots x_n^{m_k}yx_n^\omega,
\end{equation*}
by Lemma~\ref{lem-KOA1}, which is the supremum of all infinite
products over $X$  containing $x_1,\ldots,x_{n-1}$ 
only a finite number of times. 

Also, using the induction hypothesis
and \Ax4, 
\begin{align*}
(x_n^*x_1 \vee \cdots \vee x_n^*x_{n-1})^\omega
&= 
\bigvee_{1 \leq i_1,i_2,\ldots \leq n-1}x_n^*x_{i_1}x_n^*x_{i_2}\cdots\\
&=
\bigvee_{1 \leq i_1,i_2,\ldots \leq n-1}\bigvee_{k_0,k_1,\ldots}
x_n^{k_0}x_{i_1}x_n^{k_1}x_{i_2}\cdots
\end{align*}
which is the supremum of all infinite products \
over $X$ containing one of $x_1,\ldots,x_{n-1}$ 
an infinite number of times. 
Thus, $x^\omega$ is the supremum of all infinite 
products over $X$ as claimed.
The proof is complete.
\eop

\subsection{Relation to bi-inductive semiring-semimodule pairs}

Recall that when $P$ is a partially ordered set and $f$ is a function $P \to P$,
then a \emph{pre-fixed point} of $f$ is an element $x$ of $P$ with 
$xf \leq x$. Similarly, $x \in P$ is a \emph{post-fixed point} of 
$f$ if $x \leq xf$. Suppose that $f$ is monotone and has $x$ 
as its least pre-fixed point. Then $x$ is a \emph{fixed point}, i.e., 
$xf = x$, and thus the least fixed point of $f$.
Similarly, when $f$ is monotone, then the greatest post-fixed point of 
$f$,   whenever it exists, is the greatest fixed point of $f$.

When $S$ is a $^*$-continuous Kleene algebra, then $S$ 
is a Kleene algebra as defined in \cite{Kozen}.  
Thus, for all $x,y \in S$, $x^*y$ is the least pre-fixed point (and thus
the least fixed point) of the function $S \to S$ defined by 
$z \mapsto xz \vee y$ for all $z \in S$. Moreover, $yx^*$ is the least pre-fixed point 
and the least fixed point of the function $S \to S$ defined 
by $z \mapsto zx \vee y$, for all $z \in S$.

Similarly, when $(S,V)$ is a generalized $^*$-continuous 
Kleene algebra, then for all $x \in S$ and 
$v\in V$, $x^*v$ is the least pre-fixed point and the least fixed point 
of the function $V \to V$ defined by $z \mapsto xz \vee v$, where $z$
ranges over $V$. 

A \emph{bi-inductive semiring-semimodule pair} is defined as a
semiring-semimodule pair $(S,V)$ for which both $S$ and $V$ are
partially ordered by the natural order relation $\leq$ such that the
semiring and semimodule operations and the action are monotone, and
which is equipped with a star operation $^*: S \to S$ and an omega
operation $^\omega: S \to V$ such that the following hold for all
$x,y\in S$ and $v \in V$:
\begin{itemize}
\item $x^*y$ is the least pre-fixed point of the function $S \to S$ mapping 
each $z \in S$ to $xz \vee y$,
\item $x^*v$ is the least pre-fixed point of the function $V \to V$ mapping 
each $z \in V$ to $xz \vee v$,
\item $x^\omega \vee x^*v$ is the greatest post-fixed point of the function 
$V \to V$ mapping each $z \in V$ to $xz \vee v$.
\end{itemize}
A bi-inductive semiring-semimodule pair is symmetric if for all $x,y\in S$,
$yx^*$ is the least pre-fixed point of the functions $S \to S$ defined 
by $z \mapsto zx + y$ for all $z\in S$.

By the above remarks we have: 

\begin{proposition}
Suppose that $(S,V)$ is a finitary $^*$-continuous Kleene $\omega$-algebra.
When for all $x \in S$ and $v\in V$, $x^\omega \vee x^*v$ is the greatest
post-fixed point of the function $V \to V$ defined by
$z \mapsto xz \vee v$,  then $(S,V)$ is a 
symmetric bi-inductive semiring-semimodule pair.
\end{proposition}

(It is known that every bi-inductive semiring-semimodule pair is an iteration 
semiring-semimodule pair, see \cite{EsikKuich}.)

\section{B\"uchi automata in $^*$-continuous Kleene $\omega$-algebras}
\label{se:buchi}

A generic definition of B\"uchi automata in Conway semiring-semimodule
pairs was given in \cite{BEbook,EsikKuich1and2}.  In this section, we
recall this general definition and apply it to (finitary) $^*$-continuous
Kleene $\omega$-algebras. We give two different definitions of the
behavior of a B\"uchi automaton, an algebraic and a combinatorial, and
show that these two definitions are equivalent.

Suppose that $(S,V)$ is a Conway semiring-semimodule pair, $S_0$ is a
subsemiring of $S$ closed under star, and $A$ is a subset of $S$. We
write $S_0\langle A\rangle$ to denote the set of all finite sums
$s_1a_1+ \cdots + s_ma_m$ with $s_i\in S_0$ and $a_i \in A$, for each
$i = 1,\ldots,m$.

We define a (weighted) B\"uchi automaton over $(S_0,A)$ of dimension
$n \geq 1$ in $(S,V)$ as a system ${\bf A} = (\alpha, M, k)$ where
$\alpha \in S_0^{1 \times n}$ is the initial vector, $M \in S_0\langle
A \rangle ^{n \times n}$ is the transition matrix, and $k$ is an
integer $0 \leq k \leq n$.  In order to define the behavior $|{\bf
  A}|$ of ${\bf A}$, let us split $M$ into 4 parts as above, $M =
\left( \begin{smallmatrix} a & b \\
  c & d
\end{smallmatrix}\right)$, with $a \in S_0\langle A \rangle ^{k \times
k}$ the top-left $k$-by-$k$ submatrix.  Then we define
\begin{equation*}
  |\bf A| = \alpha \begin{pmatrix} (a+ bd^*c)^\omega\\ 
      d^*c(a +bd^*c)^\omega \end{pmatrix}.
\end{equation*}

We give another more combinatorial definition. A B\"uchi automaton ${\bf A} = (\alpha, M,k)$ 
of dimension $n$ may be represented as a transition system 
whose set of states is $\{1,\ldots,n\}$. For any pair of states $i,j$,
the transitions from $i$ to $j$ are determined by the $(i,j)$th entry 
$M_{i,j}$ of the transition matrix. Let $M_{i,j}  = s_1a_1+ \cdots s_ma_m$, say.
Then either there are $m$ transitions from $i$ to $j$, respectively labeled $s_1a_1, \ldots,s_na_n$,
or there is just one transition, whose label is $M_{i,j}$. 
A run of the B\"uchi automaton starting in state $i$ is an infinite path starting 
in state $i$ which infinitely often visits at least one of the first $k$ states, and the weight 
of such a run is the infinite product of the path labels. The behavior of the automaton
in state $i$ is the supremum of the weights of all runs starting in state $i$. 
Finally, the behavior of the automaton is $\alpha_1 w_1 + \cdots + \alpha_n w_n$,
where for each $i$, $\alpha_i$ is the $i$th component of $\alpha$ and $w_i$ 
is the behavior in state $i$. Let $|{\bf A}|'$ denote the behavior of ${\bf A}$ according to 
this second definition. 

\begin{theorem}
\label{thm-Buchi}
For every B\"uchi automaton ${\bf A}$ over $(S_0,A)$ in a 
finitary $^*$-continuous Kleene $\omega$-algebra, it holds that $|{\bf A}|= |{\bf A}|'$. 
\end{theorem} 

\proof This holds by Lemma~\ref{lem-MAT1} and the fact that 
any matrix semiring over a $^*$-continuous Kleene algebra is 
itself a $^*$-continuous Kleene algebra.  \eop

For completeness we also mention a Kleene theorem for the B\"uchi
automata introduced above, which is a direct consequence of the 
Kleene theorem for Conway semiring-semimodule pairs, cf. \cite{EsikKuich1and2,EsikKuichhandbook}.

\begin{theorem}
\label{thm-Kleene} 
Suppose that $(S,V)$ is a $^*$-continuous Kleene $\omega$-algebra, $S_0$ is a 
subsemiring of $S$ closed under star, and $A \subseteq S$. Then an element of $V$
is the behavior of a B\"uchi automaton over $(S_0,V)$ iff it is regular (or rational)
over $(S_0,A)$, i.e., when it can be generated from the
elements of $S_0 \cup A$ by the semiring and semimodule operations, the action,
and the star and omega operations. 
\end{theorem} 

It is a routine matter to show that an element of $V$ is rational over $(S_0,A)$ 
iff it can be written as $\bigvee_{i =1}^n x_iy_i^\omega$, where each $x_i$
and $y_i$ can be generated from $S_0 \cup A$ by $\bigvee,\cdot$ and $^*$.

\section{$^*$-continuous (generalized) Kleene algebras of locally
  finite and $\top$-continuous functions}

In the following two sections our aim is to provide additional examples
of $^*$-continuous Kleene $\omega$-algebras. These examples are 
motivated by energy problems for hybrid systems and stem from
\cite{Esiketalenergy}. We will deal with a generalization of the notion
of \emph{energy function} used in \cite{Esiketalenergy}. 
 
In this section, we define locally finite and $\top$-continuous
functions over complete lattices and show that they give rise to
(generalized) $^*$-continuous Kleene algebras.

Suppose that $L = (L,\leq)$ is a complete lattice with bottom and top
elements $\bot$ and $\top$. Then a \emph{finitely additive function}
over $L$ is a function $f: L \to L$ with $\bot f = \bot$ and $(x\vee
y)f = xf \vee yf$ for all $x,y\in L$. (We write function application
and composition in the diagrammatic order, from left to write.)  In
the same way, one may define finitely additive functions $L \to L'$
for complete lattices $L,L'$. Note that when $f: L \to L'$ is finitely
additive, then $(\bigvee X)f = \bigvee Xf$ for all finite sets $X
\subseteq L$.

Consider the collection $\FinAdd_{L,L'}$ of all finitely additive 
functions $f:L \to L'$, ordered 
pointwise. Since the (pointwise) supremum of any set of finitely additive 
functions is finitely additive, $\FinAdd_{L,L'}$ is also a complete 
lattice, in which the supremum of any set of functions 
can be constructed pointwise. The least and greatest elements are 
the constant functions with value $\bot$ and $\top$,
respectively. By an abuse of notation, we will denote these functions 
by $\bot$ and $\top$ as well. 

Suppose now that $L = L'$, so that we just write $\FinAdd_{L}$.
Since the composition of finitely additive functions is finitely additive
and the identity function $\id$ over $L$ is finitely additive, 
and since composition of finitely additive functions distributes over 
finite suprema, $\FinAdd_L$, equipped with the operation
$\vee$ (binary supremum), $;$ (composition), and the constant function 
$\bot$ and the identity function 
$\id$ as $1$, is an idempotent semiring.
It follows that when $f$ is finitely additive, 
then so is $f^* = \bigvee_{n \geq 0} f^n$. 
Moreover, $f \leq f^*$ and $f^* \leq g^*$ whenever $f \leq g$. 
Below we will usually write just $fg$ for the composition $f;g$.

All functions between complete lattices considered in this paper will
be (at least) finitely additive.

\begin{lemma}
\label{lem-rightcont}
Let $L$ be a complete lattice
and $S$ be any subsemiring of $\FinAdd_L$ closed under the star 
operation. Then $S$ is a $^*$-continuous Kleene algebra iff
for all $g,h\in S$, $g^*h  = \bigvee_{n \geq 0} g^n h$.
\end{lemma}

\proof
Suppose that the above condition holds. 
We need to show that $$f(\bigvee_{n \geq 0}  g^n) h = \bigvee_{n \geq 0} fg^n h$$
for all $f,g,h\in S$. But $f(\bigvee_{n \geq 0}  g^n) h = 
f (\bigvee_{n \geq 0} g^n h )$ 
by assumption, and we conclude that 
$f (\bigvee_{n \geq 0} g^n h ) = \bigvee_{n \geq 0} f g^n h$
since the supremum is pointwise.
\eop

\begin{definition}
  Let $L$ be a complete lattice and $f: L \to L$ a finitely additive
  function.  We call $f$ {\em locally finite} if for each $x\in L$,
  either $x f^*= \top$ or $xf^* = x \vee \cdots \vee xf^n$ for some $n
  \geq 0$.  We call $f$ \emph{$\top$-continuous} if either $f$ is the
  function $\bot$, or whenever $X\subseteq L$ with $\bigvee X = \top$,
  then $\bigvee Xf = \top$.
\end{definition}

Note that if $f: L \to L$ is a $\top$-continuous function 
which is different from the function $\bot$,  
then $\top f = \top$. 

\begin{example}
Suppose that $L$ is a complete lattice 
and $f: L \to L$ is finitely additive and 
satisfies that for each $x\in L$, 
either $xf^n = \bot$ for some $n \geq 0$, or $xf = x$, 
or $\bigvee_{n \geq 0} xf^n = \top$.
Then $f$ is locally finite. 
The functions $\id$ and $\bot$ are locally finite 
and $\top$-continuous.

When $L$ has at least two elements
and $\top$ is the supremum of all elements different from $\top$,   
then the locally finite function $f: L \to L$ with $xf = \bot$ for all $x < \top$ 
and $\top f = \top$ is not $\top$-continuous.
\end{example}

\begin{lemma}
\label{lem-finsum}
Suppose that $L$ is a complete lattice and $f: L \to L$ is finitely additive. Then the 
following conditions are equivalent.
\begin{itemize}
\item
$f$ is locally finite.
\item
For each $x\in L$, either $\bigvee_{n \geq 0} xf^n = \top$, or 
there is some $n$ with $x \vee \cdots \vee xf^n = x \vee \cdots \vee xf^{n+1}$.
\item 
For each $x\in L$, either $\bigvee_{n \geq 0} xf^n = \top$, or there is some $n$ such that $x \vee \cdots \vee xf^n = x \vee \cdots \vee xf^{m}$ for all $m \geq n$.
\item 
For each $x\in L$, either $\bigvee_{n \geq 0} xf^n = \top$, or there is some $n$ such that $x \vee \cdots \vee xf^n = x \vee \cdots \vee x f^n \vee xf^{m}$ for all $m \geq n$.
\end{itemize}
\end{lemma} 

\proof
Suppose that $x \in L$ and $xf^* < \top$, i.e., $\bigvee_{n \geq 0} xf^n < \top$. 
If $f$ is locally finite then there is some $n$ with $xf^* = x \vee \cdots \vee xf^n$.
Thus, for all $m\geq n$, 
 \begin{align*}
x \vee \cdots \vee xf^n 
&\leq 
x \vee \cdots xf^n \vee xf^m \\
&\leq  
x \vee \cdots \vee xf^m\\
&\leq  xf^*\\
&= x \vee \cdots \vee xf^n.
\end{align*}
Also, if $x\vee \cdots \vee xf^n = x\vee \cdots \vee xf^{n+1}$, 
then it follows by induction that $x\vee \cdots \vee xf^n = x\vee \cdots \vee xf^{m}$
for all $m \geq n$ and thus $xf^* = x\vee \cdots \vee xf^n$. 
Indeed, this is clear when $m = n,n+1$. Suppose that $m > n+1$. 
Then, using the induction hypothesis and the fact that $f$ is finitely additive,
\begin{align*}
x \vee \cdots \vee xf^m 
&=
x \vee (x \vee \cdots \vee xf^{m-1})f\\
&=
x \vee (x \vee \cdots \vee xf^n)f\\
&=
x \vee \cdots \vee xf^{n+1}\\
&= 
x \vee \cdots \vee xf^n. 
\end{align*}
\eop 

It is clear that the composition of $\top$-continuous 
functions is also $\top$-continuous. However, as the next 
example demonstrates, the composition of locally finite 
(and $\top$-continuous) functions is not necessarily 
locally finite. 

\begin{example}
Let $L$ be the following complete lattice (the linear sum of three
infinite chains):
$$\bot < x_0 < x_1 < \cdots < y_0 < y_1 < \cdots < z_0 < z_1 < \cdots \top$$
Since $L$ is a chain, a function $L \to L$ is finitely additive 
iff it is monotone and preserves $\bot$.

Let $f,g: L \to L$ be the following functions. First, $\bot f = \bot g = \bot$
and $\top f = \top g = \top$. Moreover, $x_i f = y_i$, $y_i f = z_i g = \top$
and $x_i g = \bot$, $y_i g = x_{i+1}$, $z_i g = \top$ for all $i$. 
Then $f,g$ are monotone, $uf^* = u \vee uf \vee uf^2$ and $ug^* = u \vee ug$ 
for all $u \in L$. Also, $f$ and $g$ are $\top$-continuous, 
since if $\bigvee X = \top$ then either 
$\top \in X$ or $X \cap \{z_0,z_1,\ldots\}$
is infinite, but then $\bigvee Xf = \bigvee Xg = \top$. 
However, $fg$ is not locally finite, since
$x_0(fg)^* = x_0\vee x_0(fg) \vee x_0(fg)^2 \cdots = x_0 \vee x_1 \vee x_2 \vee \cdots = y_0$.
\end{example}

We now show that if a set of locally finite and $\top$-continuous functions
enjoys appropriate closure properties, then it is necessarily a $^*$-continuous 
Kleene algebra. 

\begin{proposition}
  \label{prop-energy1}
  Suppose that $S \subseteq \FinAdd_L$ is closed under the operations
  $\vee,$ composition and star and contains the functions $\bot$ and
  $\id$.  Moreover, suppose that each $f \in S$ is locally finite and
  $\top$-continuous. Then $S$ is a $^*$-continuous Kleene algebra.
\end{proposition}

\proof
Suppose that $g,h\in S$. By Lemma~\ref{lem-rightcont}, it
suffices to show that 
$g^*h  = \bigvee_{n \geq 0}g^n h$.
Since this is clear when $h = \bot$, we assume that $h$ is a function
different from $\bot$.
Since $g^n h \leq g^* h$ for all $n \geq 0$, it holds that 
$\bigvee_{n \geq 0}g^n h \leq g^*h$. To prove the 
opposite inequality, suppose that $x \in L$. If 
$xg^* = \top$, then $\bigvee_{n \geq 0} xg^n = \top$, so that 
$\bigvee_{n \geq 0} xg^nh = \top$ by $\top$-continuity. 
Thus, $xg^*h = \top = \bigvee_{n \geq 0} xg^nh$. 

Suppose that $xg^* < \top$. Then there is some $m \geq 0$ with 
\begin{align*}
xg^*h 
&= (x\vee \cdots \vee xg^m)h \\
&= xh \vee \cdots \vee xg^mh\\
&\leq \bigvee_{n \geq 0}xg^nh\\
&= x(\bigvee_{n \geq 0} g^n h).
\end{align*}
The proof is complete. 
\eop


Next we study the closure properties of finitely additive $\top$-continuous 
functions in more detail. We will give a sufficient condition to the 
effect that the supremum of two such functions be also locally finite and 
$\top$-continuous.  We will also prove that if $f$ is locally finite 
and $\top$-continuous, then so is $f^*$. 

The following two propositions are also of independent interest.  

\begin{proposition}
Suppose that $L$ is a complete lattice and
$f,g_1,\ldots,g_k: L \to L$, $k \geq 0$ are 
locally finite and $\top$-continuous functions. Then 
\begin{equation*}
f^*g_1\cdots g_k = \bigvee_{n \geq 0}f^ng_1\cdots g_k.
\end{equation*}
\end{proposition}

\proof
When $k = 0$ this is obvious by definition of star, so we assume that $k > 0$.  
It is clear that $\bigvee_{n \geq 0}f^ng_1\cdots g_k \leq f^*g_1\cdots g_k$
in the pointwise order. To complete the proof, we show that 
$xf^*g_1\cdots g_k \leq \bigvee_{n \geq 0} xf^ng_1\cdots g_k$
for all $x \in L$. We may assume that 
the $g_i$ are different from the function $\bot$ since 
in the opposite case $xf^*g_1\cdots g_k = \bot$. 

Suppose first that $xf^* < \top$.
Then there is some $n$ such that 
$xf^* = x\vee \cdots \vee xf^n$.
Since the $g_i$ are finitely additive, 
it follows that $xf^* g_1\cdots g_k = 
\bigvee_{i = 0}^n xf^ig_1\cdots g_k 
\leq \bigvee_{n \geq 0} xf^ng_1\cdots g_k.$

Suppose next that $xf^* = \top$, so that $xf^*g_1\cdots g_k = \top$.
Our task is to prove that $\bigvee_{n \geq 0} xf^ng_1\cdots g_k = \top$.
To this end, let $h$ denote the function $h = g_1\cdots g_k$. 
Since each $g_i$ is $\top$-continuous, so is $h$.
Moreover, $h$ is not the function $\bot$. 
Thus, since $\bigvee \{xf^n : n \geq 0\} = \top$ and $h$ is $\top$-continuous,
\begin{equation*}
\bigvee_{n \geq 0} xf^ng_1\cdots g_k 
=
\bigvee_{n \geq 0} xf^n h = \top. \makebox[0pt][l]{\qquad \eop}
\end{equation*}

\begin{proposition}
\label{prop-fveeg}
Suppose that $L$ is a complete lattice and $f,g\in \FinAdd_L$ 
such that $f$ and $f^*g$ are locally finite and $\top$-continuous functions. 
Then 
\begin{equation*}
(f \vee g)^* = (f^*g)^*f^*.
\end{equation*}
\end{proposition}

\proof
It is clear that $(f \vee g)^n \leq \bigvee_{i\leq n}(f^*g)^i f^*$ for all $n \geq 0$,
hence $(f \vee g)^* \leq \bigvee_{n \geq 0} (f^*g)^nf^* \leq (f^*g)^*f^*$. 
Below we will prove that $$(f^*g)^*f^* \leq \bigvee_{n \geq 0} \bigvee_{k_i \geq 0}f^{k_0}g\cdots gf^{k_n}
= (f \vee g)^*.$$ In our argument, we will make use of two observations.

{\em Claim 1.}\ Suppose that $x(f^*g)^n < \top$ for some $n \geq 0$. Then there exist 
integers $k_1,\ldots,k_n\geq 0$ with 
$$ x(f^*g)^n \leq \bigvee\{ xf^{i_1}g \ldots f^{i_n}g : i_1 \leq k_1,\ldots,i_n \leq k_n\}.$$

{\em Claim 2.}\ Suppose that $x(f^*g)^nf^* < \top$ for some $n \geq 0$. Then there exist integers
$k_1,\ldots,k_n,k_{n+1}\geq 0$ with 
$$ x(f^*g)^nf^* \leq \bigvee\{ xf^{i_1}g \ldots f^{i_n}gf^{i_{n+1}} : i_1 \leq k_1,\ldots,i_{n+1} \leq k_{n+1}\}.$$

Since the proofs of these claims are similar, we only prove the first one. 
We argue by induction on $n$. When $n = 0$, our claim is clear. Suppose
now that $n > 0$. Since $x(f^*g)^n < \top$ and $f^*g$ preserves $\top$,
$x(f^*g)^{n-1} = y < \top$. Since $f$ is locally finite, 
$y f^* = y \vee \cdots \vee yf^m$ for some $m \geq 0$. 
Since $g$ is finitely additive, $yf^*g = yg \vee \cdots \vee yf^mg$.
By the induction hypothesis, 
$y \leq \bigvee\{ xf^{i_1}g \ldots f^{i_{n-1}}g : i_1 \leq k_1,\ldots,i_{n-1} \leq k_{n-1}\}$
for some $k_1,\ldots,k_{n-1}\geq 0$. Thus, using the fact that 
$f$ and $g$ are finitely additive, it follows that 
$$ x(f^*g)^n \leq \bigvee\{ xf^{i_1}g \ldots f^{i_n}g : i_1 \leq k_1,\ldots,i_n \leq k_n\}$$
where $k_n = m$.

We now return to the proof of the Proposition. Let $x \in L$. 
We want to show that $x(f^*g)^*f^* \leq x(f \vee g)^*$,
which will follow if we can prove that 
$x(f^*g)^*f^*$ is the supremum of a finite or infinite 
number of elements of the form $xf^{k_0}gf^{k_1}\cdots gf^{k_n}$
where $n \geq 0$ and $k_0,\ldots,k_n \geq 0$.
Let us denote by $\Delta_x$ the set of all such elements.

When $x = \top$, then our claim is clear. 
So we may assume that $x < \top$. We consider several cases. 

{\em Case 1.} $x(f^*g)^*f^* < \top$. Then, since $f^*g$ is
locally finite and $f^*$ is finitely additive,  
\begin{align*}
x(f^*g)^*f^* &= (x\vee \cdots \vee x(f^*g)^n)f^*\\
&= xf^*\vee \cdots \vee x(f^*g)^nf^*
\end{align*}
for some $n \geq 0$. 
Clearly, $x(f^*g)^i < \top$ and $x(f^*g)^if^* < \top$ for 
all $i \leq n$. 
Now using Claim 2 and the fact that $f$ is finitely additive,
it follows that each $x(f^*g)^i f^*$ is the supremum 
of a finite number of terms of $\Delta_x$.

{\em Case 2.}  $x(f^*g)^*f^* = \top$.

Within this case, we consider sub-cases.

{\em Sub-case 2.1.} $x(f^*g)^* < \top$. In this case, it follows 
as in Case 1 using Claim 1  
that $y = x(f^*g)^*$ is the supremum of a finite number of 
elements of $\Delta_x$.
Since $yf^* = y \vee yf \vee \cdots = \top$ and for each $k\geq 0$, 
$yf^k$ is also a finite supremum of elements of $\Delta_x$,
since $f$ is finitely additive, 
it follows that $x(f^*g)^*f^* = yf^*$ 
is a possibly infinite supremum of elements of $\Delta_x$.

{\em Sub-case 2.2.} $x(f^*g)^* = \bigvee_{n \geq 0} x(f^*g)^n = \top$.  
Suppose that $\bigvee_{i \leq n} (f^*g)^i < \top $
for all $n\geq 0$. Then, as shown above,  for each $n \geq 0$ we can write 
$\bigvee_{i \leq n} (f^*g)^i$ as a finite supremum of elements of $\Delta_x$.
Thus, $x(f^*g)^*f^* = \top = x(f^*g)^*$ is a possibly infinite supremum of elements 
of $\Delta_x$.
Suppose now that there is some
$n$ with $\bigvee_{i \leq n} (f^*g)^i = \top$. Then there is a least such $n$.
Let us denote it by $n_0$. Since $x \neq \top$, it holds $n_0 > 0$.  
Then $x(f^*g)^i < \top$ for all $i < n_0$. If also $x(f^*g)^{n_0} < \top$,
then $x(f^*g)^*f^* = \top = \bigvee_{i \leq n_0} (f^*g)^i$ is a finite supremum of 
elements of $\Delta_x$. Last, assume that $x(f^*g)^{n_0} = \top$.
Let us denote $x(f^*g)^{n_0-1}$ by $y$, so that $y < \top$ and $yf^*g = \top$. 
Note that $y$ is a finite supremum of elements of $\Delta_x$. 

Suppose that $yf^* = y \vee yf \vee \cdots = \top$. Since $f$ is finitely additive, each
$yf^i$ is a finite supremum over $\Delta_x$, and we conclude that 
so is $x(f^*g)^*f^* = \top = yf^*$ is a possibly infinite supremum over $\Delta_x$. 
If $yf^* < \top$, then $yf^* = y \vee \cdots \vee yf^n$,
and by finite additivity, we conclude that $yf^*g$ is a finite supremum of elements
of $\Delta_x$. 
\eop

Next we prove that the star of a locally finite $\top$-continuous function
also has these properties.

\begin{lemma}
\label{lem-star}
Suppose that $L$ is a complete lattice and $f\in \FinAdd_L$ 
is locally finite. Then $f^*$ is also locally finite.
And if $f$ is additionally $\top$-continuous, then so is 
$f^*$.
\end{lemma}

\proof
Suppose that $f$ is locally finite. 
We prove that $xf^{**} = x \vee xf^* = xf^*$ for all $x\in L$.
Indeed, this is clear when $xf^* =\top$, since $f^*\leq f^{**}$.
 Otherwise 
$xf^* = \bigvee_{k \leq n}xf^k$ for some $n \geq 0$.

By finite additivity, it follows that $xf^*f^* =
\bigvee_{k \leq n} xf^kf^*$. But for each $k$,
$xf^kf^* = xf^k \vee xf^{k+1} \vee \cdots \leq xf^*$
and thus $xf^* = xf^*f^*$ and $xf^*=xf^{**}$.
It follows by Lemma~\ref{lem-finsum} that $f^*$ is locally finite. 

Suppose now that $f$ is additionally $\top$-continuous.
We need to show is that $f^*$ is also
$\top$-continuous. To this end, let $X \subseteq L$ with $\bigvee X = \top$. 
Since $x \leq xf^*$, for all $x \in X$,
it holds that $\bigvee Xf^* \geq \bigvee X = \top$.
Thus $\bigvee Xf^* = \top$. 
\eop

\begin{lemma}
Suppose that $L$ is a complete lattice and $f,g,f^*g: L \to L$ are 
locally finite $\top$-continuous functions.
Then $f \vee g$ is also locally finite and $\top$-continuous. 
\end{lemma}

\proof
We know that $f \vee g$ is finitely additive. By Proposition~\ref{prop-fveeg}, 
$(f \vee g)^* = (f^*g)^*f^*$.

Suppose that there is $x \in L$ with $x(f\vee g)^* = x(f^*g)^*f^* < \top$. 
Then, since $f^*g$ and $f$ are locally finite and $\top$-continuous, we have that 
$x(f^*g)^* <\top$, $y = x(f^*g)^* = x \vee \cdots \vee x(f^*g)^n$, 
and $yf^* = y \vee \cdots \vee yf^m$ for some $n,m\geq 0$. Moreover, as shown 
in the proof of Proposition~\ref{prop-fveeg}, for each $i \leq n$ there exist 
$n_1,\ldots,n_i \geq 0$ with $x(f^*g)^i = \bigvee_{j_1\leq n_1,\ldots,j_i \leq n_i} xf^{j_1}g\ldots f^{j_i}g$.
It follows that $x(f \vee g)^* = x \vee \cdots \vee x(f\vee g)^k$ for some $k \geq 0$. 
We have proved that $f\vee g$ is locally finite.

Since $f$ and $g$ are $\top$-continuous, $f \vee g$ is clearly also 
$\top$-continuous. 
\eop

\subsection{Generalized $^*$-continuous Kleene algebras of finitely 
additive functions}

By the previous results, locally finite and $\top$-continuous functions provide examples of $^*$-continuous
Kleene algebras. Next we show that they also yield generalized $^*$-continuous
Kleene algebras.

Suppose that $L,L'$ are complete lattices, and consider the collection 
$\FinAdd_{L,L'}$ of all finitely additive functions $L \to L'$, ordered 
pointwise. Then, as noted above, $\FinAdd_{L,L'}$ is a complete lattice. 
Define a left action of $\FinAdd_L$ on $\FinAdd_{L,L'}$ by $fv = f;v$, 
for all $f \in \FinAdd_L$ and $v \in \FinAdd_{L,L'}$. It is a routine matter to check that $\FinAdd_{L,L'}$,
equipped with the above action, the binary supremum operation $\vee$ and 
the constant $\bot$ is an (idempotent) left $\FinAdd_L$-semimodule.

We extend the notion of $\top$-continuous functions $L \to L$ 
over a complete lattice to a more general situation in a natural way. 
Suppose that $L$ and $L'$ are complete lattices
and let $f: L \to L'$. We call $f$ 
$\top$-continuous if $f$ is finitely additive and 
$\bigvee Xf = \top$ whenever $X \subseteq L$ 
with $\bigvee X = \top$.

We will make use of a general fact
which is similar to Lemma~\ref{lem-rightcont}.

\begin{lemma}
\label{lem-rightcont2}
Suppose that $L,L'$ are complete lattices,
$S$ is a $^*$-continuous Kleene algebra in $\FinAdd_L$, 
and $V \subseteq \FinAdd_{L,L'}$ is an $S$-semimodule. 
Then 
$(S,V)$ is a generalized $^*$-continuous Kleene algebra
iff  for all $g\in S$ and $v \in V$,
$g^*v = \bigvee_{ n\ge 0} g^n v$ (and thus $\bigvee_{ n\ge 0} g^n v\in V$).
\end{lemma}

\proof
The above condition yields that for all $f,g\in S$ and $v \in V$,
$fg^*v = \bigvee_{n \geq 0}  f g^n v$, so that \Ax0 holds.
\eop

\begin{proposition}
  \label{prop-energy2}
  Let $S$ be a $^*$-continuous Kleene algebra of locally finite and
  $\top$-continuous functions in $\FinAdd_L$, and let $V$ be a
  subsemimodule of $\FinAdd_{L,L'}$ closed under the induced
  $S$-action, which contains only $\top$-continuous functions. Then
  $(S,V)$ is a generalized $^*$-continuous Kleene algebra.
\end{proposition}

\proof
The proof is similar to that of Proposition~\ref{prop-energy1}. 
\eop

\begin{remark}
  Actually it suffices to require in the above proposition the
  following. For all $v\in V$, $v \neq \bot$, and for all $f \in S$
  and $x \in L$, if $xf^* = \top$ then $\bigvee_{n \geq 0}xf^nv =
  \top$.  Then $(S,V)$ is a generalized $^*$-continuous Kleene
  algebra.  Indeed, suppose that $f\in S$ and $v \in V$. In order to
  prove that $f^*v = \bigvee_{n \geq 0}f^nv$, it suffices to show that
  $xf^*v \leq \bigvee_{n \geq 0} xf^nv$ for all $x$.

  If $xf^* < \top$, then $xf^* = x \vee \cdots \vee xf^n$ for some
  $n$.  Since $v$ is finitely additive, $xf^*v = xv \vee \cdots \vee
  xf^n v \leq \bigvee_{n \geq 0} xf^nv$.  If $xf^* = \top$, then
  $xf^*v = \top$ unless $v$ is $\bot$.  It follows that $f^*v =
  \bigvee_{n \geq 0} f^nv$.
\end{remark}

\section{$^*$-continuous Kleene $\omega$-algebras of locally finite and 
$\top$-continuous functions}

In this section, our aim is to study $^*$-continuous Kleene
$\omega$-algebras of finitely additive functions.  For this reason, we
define an infinite product mapping $\omega$-sequences of finitely
additive functions over a complete lattice into the $2$-element
lattice $\two$.

In this section, let $L$ be an arbitrary complete lattice
and let $L' = \two$, the $2$-element lattice $\{\bot,\top\}$.
Our aim is to study generalized $^*$-continuous Kleene $\omega$-algebras $(S,V)$,
 where $S$ consists of locally finite $\top$-continuous functions 
$L \to L$ and $V$ consists of $\top$-continuous functions 
$L \to \two$.

For each sequence $f_0,f_1,\ldots \in \FinAdd_L$, define $g = \prod_{n
  \geq 0}f_n: L \to \two$ by $xg = \bot$ iff there is an index $n\ge
0$ for which $xf_0\cdots f_n = \bot$.  Thus, if $x,
xf_0,xf_0f_1,\ldots$ are all different from $\bot$, then $x\prod_{n
  \geq 0}f_n = \top$, otherwise $x\prod_{n \geq 0}f_n = \bot$.  Note
that in the latter case the sequence $x, xf_0,xf_0f_1,\ldots$ is
eventually $\bot$.

It is easy to see that $\prod_{n \geq 0}f_n$ is also finitely additive. 
Indeed, $\bot \prod_{n \geq 0}f_n = \bot$ clearly holds, and for all $x,y \in L$,
if $x \leq y $ in $L$, then $x\prod_{n \geq 0} f_n  \leq y\prod_{n \geq 0}f_n$. 
Thus, to prove that for all $x,y \in L$, it holds that $(x \vee y)\prod_{n \geq 0}f_n
= x\prod_{n \geq 0}f_n \vee y\prod_{n \geq 0}f_n$, it suffices to show that 
if $x\prod_{n \geq 0}f_n= y\prod_{n \geq 0}f_n = \bot$, then $(x\vee y)\prod_{n \geq 0}f_n = \bot$. 
But if $x\prod_{n \geq 0}f_n= y\prod_{n \geq 0}f_n= \bot$, then there exist $m,k$ 
with $xf_0\cdots f_m = yf_0\cdots f_k = \bot$. Let $\ell =\max\{m,k\}$. We have that 
$(x \vee y)f_0\cdots f_\ell  = xf_0\cdots f_\ell \vee yf_0\cdots f_\ell = \bot$, 
and thus $(x\vee y)\prod_{n\geq 0}f_n = \bot$.

It is clear that \Ax1 and \Ax2 hold.

\begin{proposition}
Suppose that $f_0,f_1,\ldots$ are finitely additive functions $L \to L$ and $0 = n_0 \leq n_1 \leq \cdots$ 
is a sequence of integers which increases without a bound. Let $g_k = f_{n_k}\cdots f_{n_{k+1} -1}$
for all $k \geq 0$, so that each $g_k$ is also finitely additive. Then
\begin{align*}
\prod_{n \geq 0}f_n &= f_0\prod_{n \geq 0}f_{n  + 1}\\
\prod_{n \geq 0} f_n &= \prod_{k \geq 0}g_k.
\end{align*}
\end{proposition}

\begin{proposition}
  The following property holds for all finitely additive functions
  $f_0,f_1,\ldots$ and $g_0,g_1,\ldots$ $L \to L$:
  \begin{equation}
    \label{lem-energy-Ax3}
    \prod_{n \geq 0}(f_n \vee g_n) =  \bigvee_{h_n\in \{f_n,g_n\}} \prod_{n \geq 0} h_n.
  \end{equation}
  In particular, \Ax3 holds.
\end{proposition}

\proof
Suppose that $f_0,g_0,f_1,g_1,\ldots $ are in $\FinAdd_L$. 
Since infinite product is monotone, the term on the right-hand side of (\ref{lem-energy-Ax3})
is less than or equal to the term on the left-hand side. To prove that equality holds, 
let $x \in L$ and suppose that $x\prod_{n \geq 0}(f_n \vee g_n) = \top$. 
It suffices to show that there is a choice of the functions $h_n\in \{f_n,g_n\}$ such that 
$x \prod_{n \geq 0} h_n = \top$. Consider the full (ordered) binary tree
where each node at level $n \geq 0$ is the source of an edge labeled $f_n$ 
and an edge labeled $g_n$, ordered as indicated. We can assign to each node $u$  
the composition $h_u$ of the functions that occur as the labels of the edges along  
the unique path from the root to that node. Let us mark a node $u$ if $xh_u \neq \bot$. 
Since $x \prod_{n \geq 0}(f_n \vee g_n) = \top $, 
each level contains a marked node. Moreover, whenever a node is marked and has a predecessor,
its predecessor is also marked. By K\"onig's lemma there is an infinite path going through marked nodes. 
This infinite path gives  rise to the sequence $h_0,h_1,\ldots$ with 
$x\prod_{n \geq 0}h_n = \top$.
\eop

\begin{proposition}
Suppose that $f,g_0,g_1,\ldots$ are locally finite and $\top$-continuous functions $L \to L$. 
Then $\prod_{n \geq 0}f^*g_n = \bigvee_{k_0,k_1,\ldots \geq 0} \prod_{n \geq 0} f^{k_n}g_n$,
i.e., \Ax4 holds. 
\end{proposition}

\proof
Since $f$ is locally finite and $\top$-continuous, so is $f^*$, by Lemma~\ref{lem-star}. 
Let $x \in L$. Suppose that $x\prod_{n \geq 0}f^*g_n = \top$. We want to prove that there exist
integers $k_0,k_1,\ldots$ such that $x\prod_{n \geq 0} f^{k_n}g_n = \top$. 
Consider the ordered infinite binary tree  and label the left out-edge of 
each vertex $v$ by $f$ and the right out-edge by $g_n$ iff the unique path 
from the root to $v$ contains exactly $n$ right out-edges. To each vertex $z$
we can canonically associate the function $h_z$ which is the composition
of the edge labels that occur along the path from the root to $z$.
Let us mark a vertex $z$ if following condition 
holds: if $z_1,\ldots,z_k$ are some ancestors of $z$ such that
each $z_{i+1}$ is the left successor of $z_i$, for $i < k$,
then it is not the case that $xh_{z_k} \leq xh_{z_1}\vee \cdots \vee xh_{z_{k-1}}$.

Since $x\prod_{n \geq 0}f^*g_n = \top$,
for each $n$ there is a marked vertex which is the target of an edge
labeled $g_n$. Moreover, any ancestor of a marked vertex is marked. 
By K\"onig's lemma, there is an infinite path going through marked vertices.
If this path contains infinitely many right edges, we are done, 
since this infinite path determines the integers $k_n$, $n \geq 0$.
Indeed, the edge labels of the path form an infinite word 
$f^{k_0}g_0f^{k_1}g_1\cdots$.

Suppose the path contains only a finite number of right edges.
For simplicity, we may as well assume that the path contains only
left edges. This means that $x,xf,xf^2,\ldots$ satisfy that for
all $n$, $xf^n \not \leq \bigvee_{i < n} xf^i$ 
and thus $xf^* = \top$,
since $f$ is locally finite and $\top$-continuous. 
Since none of the functions $g_0,g_1,\ldots$ is $\bot$,
 $v = \prod_{n \geq 0}g_n$ is not $\bot$,
since $\top v = \top$. Now 
$\bigvee_{n \geq 0} f^nv = f^*v$ and since $xf^*v = \top$, also 
$\bigvee_{n \geq 0} xf^nv =\top$. Thus, $xf^mv \neq \bot$ for some $m$. 
We may thus choose $k_0 = m$ and $k_n = 0$ for all $n > 0$.
\eop 

\begin{corollary}
  \label{co:genkleene->scontkleeneom}
  Suppose that $L$ is a complete lattice and $(S,V)$ is a generalized
  $^*$-continuous Kleene algebra of locally finite and
  $\top$-continuous functions $L \to L$ and $\top$-continuous
  functions $L \to \two$. Suppose that $\prod_{n \geq 0}f_n \in V$ for
  all sequences $f_0,f_1,\ldots$ of functions in $S$.  Then $(S,V)$ is
  a $^*$-continuous Kleene $\omega$-algebra.
\end{corollary}

\begin{corollary}
  Suppose that $L$ is a complete lattice and $(S,V)$ is a generalized
  $^*$-continuous Kleene algebra of locally finite and
  $\top$-continuous functions $L \to L$ and $\top$-continuous
  functions $L \to \two$. Suppose that $\prod_{n \geq 0}f_n \in V$ for
  all finitary sequences $f_0,f_1,\ldots$ of functions in $S$.  Then
  $(S,V)$ is a finitary $^*$-continuous Kleene $\omega$-algebra.
\end{corollary}

Suppose that $L$ is a complete lattice and $f_0,f_1,\ldots$ is a
sequence of locally finite $\top$-continuous functions $L \to L$, and
consider the function $v = \prod_{n \geq 0}f_n : L \to \two$ defined
above. While $v$ is finitely additive, it may not be
$\top$-continuous. Below we provide a sufficient condition under which
this function is also $\top$-continuous.

\begin{lemma}
  \label{lem-infprodenergy}
  Suppose that $L$ is a complete lattice for which it holds that
  whenever $\bigvee X = \top$ for some $X \subseteq L$, then for all
  $x < \top$ in $L$ there is $y\in X$ with $x \leq y$.  If
  $f_0,f_1,\ldots$ is a sequence of locally finite $\top$-continuous
  functions $L\to L$, then $\prod_{n\geq 0}f_n$ is $\top$-continuous.
\end{lemma}

\proof
Let $v = \prod_{n \geq 0}f_n$. We already know that $v$ is finitely additive.
We need to show that if $v \neq \bot$, then $v$ is $\top$-continuous.
But if $v\neq \bot$, then there is some $x < \top$ with $xv = \top$,
i.e., such that $xf_0\cdots f_n > \bot$ for all $n$. By assumption,
there is some $y \in X$ with $x \leq y$. It follows that 
$yf_0\cdots f_n \geq xf_0\cdots f_n > \bot$ for all $n$ 
and thus $\bigvee Xv = \top$.
\eop

\subsection{Fixed points}

In this short section, we relate $^*$-continuous Kleene $\omega$-algebras 
of locally finite and $\top$-continuous functions to bi-inductive 
semiring-semimodule pairs.

\begin{lemma} 
Suppose that $L$ is a complete lattice and $(S,V)$ is a
generalized  $^*$-continuous Kleene algebra of 
locally finite $\top$-continuous functions $L \to L$ and $\top$-continuous
functions $L \to \two$.
Suppose that for each $f \in S$, $f^\omega = \prod_{n \geq 0} f \in V$.
Then for all  $f\in S$ and $v \in V$, $f^\omega \vee f^*v$ is the greatest 
post-fixed point of the map $V \to V$ given by $z \mapsto fz \vee v$.
\end{lemma}

\proof Since $ff^\omega = f^\omega$, we have that 
$f(f^\omega \vee f^*v)\vee v  = ff^\omega \vee ff^*v \vee v = f^\omega \vee f^*v$.

Suppose that $u \in V$ with $u \leq fu \vee v$. Then 
$u \leq f^nu \vee f^{n-1}v \vee \cdots \vee v$ for all $n \geq 0$.
We want to show that $xu \leq xf^\omega \vee xf^*v$ for all $x \in L$.
This is clear when $xf^\omega = \top$. So suppose that $xf^\omega = \bot$.
Then $xf^n = \bot$ for some $n$. Thus, 
$xu \leq xf^{n-1}v \vee \cdots xv \leq xf^*v = xf^\omega \vee xf^*v$. 
\eop

\begin{corollary} 
  \label{co:genstconkle->indsemim}
  Suppose that $L$ is a complete lattice and $(S,V)$ is a generalized
  $^*$-continuous Kleene algebra of locally finite $\top$-continuous
  functions $L \to L$ and $\top$-continuous functions $L \to \two$.
  Suppose that for each $f \in S$, $f^\omega = \prod_{n \geq 0} f \in
  V$.  Then $(S,V)$ is a symmetric bi-inductive semiring-semimodule
  pair.
\end{corollary}



\section{Application: Energy Problems}

We finish the paper by showing how the setting developed in the last
two sections can be applied to solve so-called energy problems.  We
refer to~\cite{Esiketalenergy} for a more detailed account of, and
motivation for, energy problems in the context of formal modeling and
verification.

Let $L=[ 0, \top]_\bot$ be the complete lattice of nonnegative real
numbers together with $\top= \infty$ and an extra bottom element
$\bot$, and extend the usual order and operations on real numbers to
$[ 0, \top]_\bot$ by declaring that $\bot< x< \top$, $\bot- x= \bot$
and $\top+ x= \top$ for all nonnegative reals $x$.

Note that $L$ satisfies the condition required in
Lemma~\ref{lem-infprodenergy}: If $X\subseteq[ 0, \top]_\bot$ is such
that $\bigvee X= \top$, then for all $x\in L$ with $x< \top$ there is
$y\in X$ for which $x\le y$.

An \emph{energy function} is a mapping $f: [ 0, \top]_\bot\to [ 0,
\top]_\bot$ for which $\bot f= \bot$, $\top f= \bot$ if $x f= \bot$
for all $x< \top$ and $\top f= \top$ otherwise, and $y f\ge x f+ y- x$
whenever $\bot< x< y< \top$.  The set of such functions is denoted
$\E$.

The above entails that for all $f\in \E$ and all $x< y\in [ 0,
\top]_\bot$, $x f= \top$ implies $y f= \top$ and $y f= \bot$ implies
$x f= \bot$.  Note that energy functions are monotone, hence finitely
additive, and that $\E$ is closed under (pointwise) binary supremum
$\vee$ and composition.

\begin{lemma}
  Energy functions are $\top$-continuous.
\end{lemma}

\proof Let $X\subseteq[ 0, \top]_\bot$ such that $\bigvee X= \top$ and
$f\in \E$, $f\ne \bot$.  We have $X\ne \{ \bot\}$, so let $x_0\in
X\setminus\{ \bot\}$ and, for all $n\ge 0$, $x_n= x_0+ n$.  Let $y_n=
x_n f$.
If $y_n= \bot$ for all $n\ge 0$, then also $n f= \bot$ for all $n\ge
0$ (as $x_n\ge n$), hence $f= \bot$.  We must thus have an index $N$
for which $y_N> \bot$.  But then $y_{ N+ k}\ge y_N+ k$ for all $k\ge
0$, hence $\bigvee X f= \top$. \eop

\begin{lemma}
  For $f\in \E$, $f^*$ is given by $x f^*= x$ if $x f\le x$ and $x
  f^*= \top$ if $x f> x$.  Hence $f$ is locally finite and $f^*\in
  \E$.
\end{lemma}

\proof We have $\bot f^*= \bot$ and $\top f^*= \top$.  Let $x\ne \bot,
\top$.  If $x f\le x$, then $x f^n\le x$ for all $n\ge 0$, so that
$x\le \bigvee_{ n\ge 0} x f^n\le x$, whence $x f^*= x$.
If $x f> x$, then let $a= x f- x> 0$.  We have $x f\ge x+ a$, hence $x
f^n\ge x+ n a$ for all $n\ge 0$, so that $x f^*= \bigvee_{ n\ge 0} x f^n=
\top$. \eop

We can now apply Proposition~\ref{prop-energy1}:

\begin{corollary}
  $\E$ is a $^*$-continuous Kleene algebra.
\end{corollary}

\begin{remark}
  It is not true that $\E$ is a continuous Kleene algebra: Let $f_n,
  g\in \E$ be defined by $x f_n= x+ 1- \frac1{ n+ 1}$ for $x\ge 0$,
  $n\ge 0$ and $x g= x$ for $x\ge 1$, $x g= \bot$ for $x< 1$.  Then
  $0( \bigvee_{ n\ge 0} f_n) g=( \bigvee_{ n\ge 0} 0 f_n) g= 1 g= 1$,
  whereas $0 \bigvee_{ n\ge 0}( f_n g)= \bigvee_{ n\ge 0}( 0 f_n g)=
  \bigvee_{ n\ge 0}(( 1- \frac1{ n+ 1}) g)= \bot$.

  Also note that not all locally finite functions $f: [ 0,
  \top]_\bot\to [ 0, \top]_\bot$ are energy functions: the function
  $f$ defined by $x f= 1$ for $x< 1$ and $x f= x$ for $x\ge 1$ is
  locally finite, but $f\notin \E$.
\end{remark}

Now let $L'= \two$, and let $\V$ denote the $\E$-semimodule of all
$\top$-continuous functions $[ 0, \top]_\bot\to \two$.  For $f_0,
f_1,\dotsc\in \E$, define an infinite product $f= \prod_{ n\ge 0}
f_n:[ 0, \top]_\bot\to \two$ by $x f= \bot$ if there is an index $n$
for which $x f_0\dotsm f_n= \bot$ and $x f= \top$ otherwise.  By
Lemma~\ref{lem-infprodenergy}, $\prod_{ n\ge 0} f_n$ is
$\top$-continuous, i.e.,~$\prod_{ n\ge 0} f_n\in \V$.

By Proposition~\ref{prop-energy2}, $( \E, \V)$ is a generalized
$^*$-continuous Kleene algebra.  We can now apply
Corollaries~\ref{co:genkleene->scontkleeneom}
and~\ref{co:genstconkle->indsemim}:




\begin{corollary}
  $( \E, \V)$ is a $^*$-continuous Kleene $\omega$-algebra and a
  symmetric bi-inductive semiring-semimodule pair.
\end{corollary}

The paper~\cite{Esiketalenergy} is concerned with reachability and
B\"uchi acceptance in so-called \emph{energy automata}, which are
finite automata labeled with energy functions.  This is useful for
solving certain energy and resource management problems in real-time
and hybrid systems, see~\cite{Esiketalenergy} for details.  

Let ${\bf A}=( \alpha, M, k)$ be an energy automaton of dimension
$n\ge 1$, with $\alpha\in\{ \bot, \id\}^{ 1\times n}\subseteq \E^{
  1\times n}$, $M\in \E\langle \E\rangle^{ n\times n}$ and $0\le k\le
n$ as in Section~\ref{se:buchi} (hence $S_0= A= \E$ in the notation of
that section).  Define $\zeta\in\{ \bot, \id\}^{ n\times 1}$ by
$\zeta_i= \id$ if $i\le k$, $\zeta_i= \bot$ for $i> k$.  We can now
apply the results of this paper to conclude the following:

\begin{theorem}
  A final state in $\bf A$ is reachable with initial energy $x\in[ 0,
  \top]_\bot$ iff $x \alpha M^* \zeta> \bot$.  There is an infinite
  run in $\bf A$ from initial energy $x$ which visits an accepting
  state infinitely often iff $x \alpha|{ \bf A}|= \top$.
\end{theorem}

\section{Conclusion}

We have introduced continuous and (finitary and non-finitary)
$^*$-continuous Kleene $\omega$-algebras and exposed some of their
basic properties.  Continuous Kleene $\omega$-algebras are idempotent
complete semiring-semimodule pairs, and conceptually, $^*$-continuous
Kleene $\omega$-algebras are a generalization of continuous Kleene
$\omega$-algebras in much the same way as $^*$-continuous Kleene
algebras are of continuous Kleene algebras: In $^*$-continuous Kleene
algebras, suprema of finite sets and of sets of powers are required to
exist and to be preserved by the product; 
in \mbox{$^*$-continuous} Kleene $\omega$-algebras these suprema are
also required to be preserved by the infinite product.

It is known that in a Kleene algebra, $^*$-continuity is precisely
what is required to be able to compute the reachability value of a
weighted automaton (or its power series) using the matrix star
operation.  Similarly, we have shown that the B\"uchi values of
automata over $^*$-continuous $\omega$-algebras can be computed using
the matrix omega operation.


We have seen that the sets of finite and infinite languages over an
alphabet are the free continuous Kleene $\omega$-algebras, and that
the free finitary $^*$-continuous Kleene $\omega$-algebras are given
by the sets of regular languages and of finite unions of finitary
infinite products of regular languages.  A characterization of the
free (non-finitary) $^*$-continuous Kleene $\omega$-algebras (and
whether they even exist) is left open.

We have shown that other examples of $^*$-continuous Kleene
$\omega$-algebras are given by locally finite and $\top$-continuous
functions over complete lattices.  We have seen that every
$^*$-continuous Kleene $\omega$-algebra is an iteration
semiring-semimodule pair, which permits to compute the behavior of
B\"uchi automata with weights in a $^*$-continuous Kleene
$\omega$-algebra using $\omega$-powers of matrices.  Hence the
algebraic setting developed here can be employed to solve energy
problems for hybrid systems.


\thebibliography{nn}

\bibitem{BerstelReutenauer}
J. Berstel and Ch.  Reutenauer, \emph{Noncommutative Rational Series With Applications},  Cambridge University Press, 2010.

\bibitem{BEbook}
S.L. Bloom and Z. \'Esik, \emph{Iteration Theories - The Equational Logic of Iterative Processes}. EATCS Monographs on Theoretical Computer Science, Springer 1993.

\bibitem{Bouyeretaltimed}
P. Bouyer, U. Fahrenberg, Kim G. Larsen, N. Markey and J. Srba,
Infinite runs in weighted timed automata with energy
constraints. FORMATS 2008, LNCS 5215, Springer, 33--47.

\bibitem{Conway} J.H. Conway, \emph{Regular Algebra and Finite Machines}, Chapman and Hall, 1971. 

\bibitem{EsItsem}
Z. \'Esik, Iteration semirings, DLT 2008, LNCS 5257, Springer, 1--21. 

\bibitem{Esiketalenergy}
Z. \'Esik, U. Fahrenberg, A. Legay and K. Quaas,
Kleene algebras and semimodules for Energy Problems. ATVA 2013, 
LNCS 8172, Springer, 102--117. 
  
\bibitem{EsikKuichrational}
Z. \'Esik and W. Kuich,  Rationally additive
semirings. J. Univ. Comput. Sci.,  8(2002), 173--183.


\bibitem{EsikKuichIND}
Z. \'Esik and W. Kuich,
Inductive $^*$-semirings, Theor. Comput. Sci., 324(2004), 3--33.

\bibitem{EsikKuich1and2}
Z. \'Esik and W. Kuich,
 A semiring-semimodule generalization of $\omega$-regular languages, Parts 1 and 2, J. Automata, Languages and Combinatorics, 10(2005), 203--242, 243--264.

\bibitem{EsikKuich}
Z. \'Esik and W. Kuich,
On iteration semiring-semimodule pairs, Semigroup Forum, 75(2007), 129--159.

\bibitem{EsikKuichhandbook}
Z. \'Esik and W. Kuich, Finite automata. In: M. Droste, W. Kuich, H. Vogler, eds., \emph{Handbook of Weighted Automata}, Springer, 2009, 69--104.


\bibitem{Golan}
J. Golan, 
\emph{Semirings and their Applications}, Springer, 1999. 


\bibitem{Kozen} D. Kozen, On Kleene Algebras and Closed Semirings,
  MFCS 1990, LNCS 452, Springer, 26--47.


\bibitem{Kozen2} D. Kozen, A Completeness Theorem for Kleene Algebras and the Algebra of Regular Events. 
Inf. Comput. 110(1994), 366--390.

\bibitem{Krob}
D. Krob, Complete Systems of B-Rational Identities. Theor. Comput. Sci. 89(1991), 207--343.


\bibitem{PerrinPin}
D. Perrin and J.-E. Pin, \emph{Infinite Words}, Pure and Applied Mathematics
vol. 141, Elsevier, 2004.

\bibitem{Salomaa}
A. Salomaa, 
Two complete axiom systems for the algebra of regular events,
J. ACM, 13(1966), 158--169.
 
\bibitem{Wilke} Th. Wilke,
An Eilenberg theorem for infinity-languages, ICALP 1991, LNCS 510, 588--589.
\end{document}